\documentclass{elsarticle}
\biboptions{sort&compress}
\usepackage{amsmath}
\usepackage{amssymb}
\usepackage{amsthm}
\usepackage[margin=0.6in, paper=a4paper]{geometry}
\usepackage[colorlinks = true]{hyperref}
\usepackage[nameinlink]{cleveref}
\usepackage{graphicx}
\graphicspath{{Figures/}}
\usepackage{subcaption}

\newcommand{\inner}[2]{\langle#1,#2\rangle}
\newcommand{\grad}{\mathop{\rm grad}\nolimits}
\newcommand{\curl}{\mathop{\rm curl}\nolimits}
\renewcommand{\div}{\mathop{\rm div}\nolimits}
\newcommand{\tr}{\mathop{\rm tr}\nolimits}
\newcommand{\R}{\mathbb{R}}
\newcommand{\newterm}[1]{\textbf{#1}}
\newcommand{\mass}{\mathsf{M}}
\newcommand{\length}{\mathsf{L}}
\renewcommand{\time}{\mathsf{T}}

\journal{arXiv}
\begin{document}
\hypersetup{
  linkcolor = blue,
  citecolor = green,
  urlcolor = cyan,
  pdfauthor = {Name}
}

\begin{frontmatter}

\title{Calculus with combinatorial differential forms for fluid flow analysis in porous and fractured media}
\author[1]{Changhao Liu\corref{fn1}}
\author[1]{Kiprian Berbatov}
\author[2]{Majid Sedighi}
\author[1]{Andrey P. Jivkov\corref{fn2}}
\address[1]{Department of Mechanical and Aerospace Engineering, The University of Manchester, Oxford Road, Manchester M13 9PL, UK}
\address[2]{Department of Civil Engineering and Management, The University of Manchester, Oxford Road, Manchester M13 9PL, UK}
\cortext[fn1]{Corresponding author: changhao.liu-2@postgrad.manchester.ac.uk}
\cortext[fn2]{Corresponding author: andrey.jivkov@manchester.ac.uk}

\begin{abstract}
The fabric of porous and fractured media contains solid regions (grains) and voids.
The space conducting fluids is a system of connected voids with variable geometries.
Relative to the grain sizes, the voids can be voluminous with three comparable large extensions, narrow expansive with two comparable large extensions and one smaller extension, and thin long with one comparable large extension and two smaller extensions.
The widely used representation of void spaces by  systems of spheres connected by cylinders (pore network models) is an acceptable approximation for some special cases, but not for most porous and fractured media.
We propose a flexible method for modelling such media by mapping their measured fabric’s characteristics – void and grain volume distributions and shapes – onto polyhedral tessellations of space.
The map assigns voluminous voids and grains to polyhedrons (3D), narrow expansive voids to some polyhedral faces (2D), and thin long voids to some polyhedral edges (1D), as dictated by experimental data.
The analysis of transport through such discrete structures with components of different dimensions is performed by a novel mathematical method, which uses combinatorial differential forms to represent physical properties and their fluxes, as well as structure-preserving operators on such forms to formulate the conservation laws exactly and directly in matrix form, ready for computation.
The method allows for individual material properties, such as conductivity, to be assigned to voids of all dimensions, so that the three types of voids are suitably represented.
Publicly available XCT images of four different rocks are used to test the method. Image analysis is used to obtain their fabrics’ characteristics and 30 model realisations with a statistically equivalent fabric are constructed for each rock. The new mathematical method is used to calculate a range of realisation-dependent permeabilities. Fluid dynamics simulations in the imaged pore spaces are performed to obtain a single, image-dependent, permeability for each rock. 
The results demonstrate the ability of the proposed method to predict the spread of transport coefficients for a given fabric using only fabric’s statistical characteristics. They agree with experimentally measured and calculated by fluid dynamics permeabilities.
The key benefits are the significantly lower computational cost compared to fluid dynamics simulations, the opportunity for up-scaling to large rock volumes with sufficiently accurate fabric’s statistics, and the potential for investigating the effect of pore space evolution on transport properties.
\end{abstract}
 

\end{frontmatter}

\section{Introduction}
\noindent Developing a method for calculating fluid flow in porous and fractured materials is a challenging task.
The classical formulation is based on the approximation of these materials as continuous media \cite{allen2021mathematics}.
In such case, a scalar physical property, fluid volume, is represented by a scalar density field, continuously distributed over the points of a material domain. The volume density is related to pressure via an equation of state. The pressure gradient is related to fluid flow velocity via a constitutive relation (Darcy’s law). The flow continuity at material points is given by the divergence of the flow velocity, leading to a partial differential equation (PDE) for volume density.
This formulation can only be used at engineering length scales where materials are sufficiently well approximated as continua.
Fluid flow can then be simulated by established numerical methods for solving PDEs, such as finite elements \cite{larsson2013finite} or finite volumes \cite{moukalled2016finite}, with macroscopic/bulk material properties.
 
The properties of real materials, however, are non-uniformly distributed; in fact, they are discontinuously distributed in porous and fractured media, e.g., between the solid phase and the pore space \cite{holzer2023tortuosity}.
The macroscopic/bulk properties are emergent, i.e., they depend on the arrangements (fabric) and interactions of internal microstructural elements \cite{nemat1993micromechanics}.
The direct approaches to obtain the bulk properties are by experiments at larger length scales and by simulations of fluid flow in the segmented pore space form X-ray Computed Tomography (XCT) images at smaller length scales.
In both cases, the values will depend on the fabric of the test pieces.
A large material domain will require multiple experiments or imaging and simulations with different test pieces to determine the range of fabric-dependent values.
The experiments and the imaging, as well as the classic simulation method, Computational Fluid Dynamics (CFD), are expensive to conduct in large numbers.
 
Bulk properties are also derived in engineering practice using approximate relations to some averaged pore space characteristics, such as porosity, tortuosity, and pore surface area \cite{costa2006permeability,sarout2012impact}.
A typical example of such a relation is the Kozeny-Carman equation \cite{carman1956flow}, which has been widely used in the porous media community and has been modified several times \cite{chen2021model,wang2022hydraulic}.
Irrespective of the way bulk properties are obtained, the return to the continuum description can only serve engineering scale simulations.
It cannot provide understanding of the effect of the fabric on the emergent behaviour, and critically the effect of fabric evolution on this behaviour.
Such understanding requires a discrete representation of the pore space that reflects as closely as possible its measured characteristics.
 
Conceptually, the simplest approach to consider the pore space discreteness, thus allowing for analysis of fabric effect on emergent behaviour, is the pore network modelling (PNM) \cite{jivkov2014network,xiong2015measurement}.
In PNM, pores are represented by spherical containers with sizes following measured pore size distribution, connected by throats represented by cylinders with sizes following measured throat size distribution.
Thus, the flow is 1-dimensional along the throats and the mass conservation is established in pores.
However, real pore spaces are rarely amenable to pore network representations, which is evidenced by markedly different pore networks constructed directly from XCT images by different characterisation algorithms \cite{baychev2019reliability}.
In fact, typical pore spaces are collections of features that appear to have different dimensions. This can be observed in the recently reported database of imaged rocks \cite{santos2022data}. For example, some pores are elongated in one direction, as cylinders/rods, while others are elongated in two directions, as plates/blades (this category includes fractures).
Relative to the 3D grains, the former appear as 1D and the latter as 2D features. The classification of pore shapes can be done by the parameter compactness related to three axial lengths of pores \cite{jahne2005digital,holzer2023tortuosity}.
Capturing this complexity requires a different approach to representation and analysis of porous and fractured media to underpin the advance in understanding the effects of fabric and its evolution on fluid flow.
Such an approach should capture the finite, discrete natures/properties of pore space components and be capable of computing its multi-dimensional interactions between internal elements.
 
The aim of this paper is to demonstrate how a new mathematical description of the fabric of porous and fractured media as a collection of discrete entities of different dimensions, together with a new calculus on such collections, provides the essential flexibility to model fabrics of arbitrary complexity and to calculate fluid flow and their macroscopic transport properties effectively and efficiently.
In \Cref{sec:description_and_analysis} we present the mathematical formalism for describing collections of discrete entities and for analysis of their behaviour.
In \Cref{sec:materials_and_models}, we provide the fabric characteristics of four rocks obtained by analyses of publicly available XCT images and show how these are mapped to the elements of collections to construct statistically representative models.
In \Cref{sec:results_and_discussion}, we show the results of fluid flow analysis with the new method and compare these with fluid dynamics simulations performed on the imaged pore spaces to demonstrate the accuracy and efficiency of the proposed approach.

\section{Description and analysis of porous media}
\label{sec:description_and_analysis}
\begin{figure}[!ht]
  \centering
  \includegraphics[scale=.6]{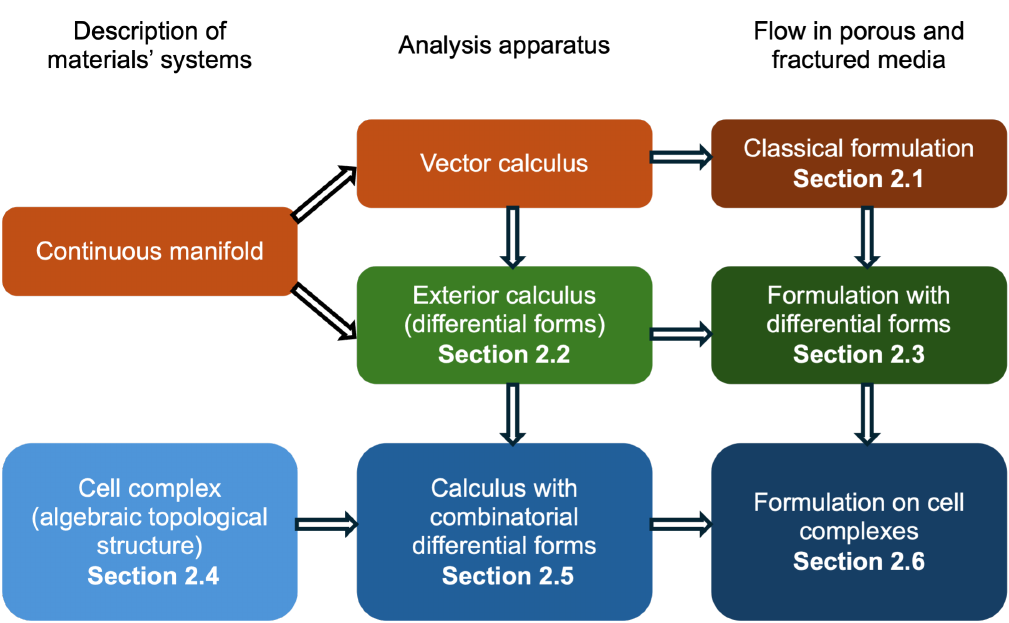}
  \caption{Pathway for development of calculus on complex materials' systems}
  \label{fig:Development}
\end{figure}

\noindent The development of a method for description  and analysis of complex materials' systems involves mathematical concepts that are not yet included in the mainstream mathematical education (exterior calculus) or are a recent development (calculus with combinatorial differential forms). Therefore, we provide an incremental introduction to these concepts as illustrated in \Cref{fig:Development}. Our aim is to arrive at the bottom right block of the diagram by building the required language and operations. It is assumed that the reader is familiar with vector calculus, so this block is not discussed.

\subsection{Classical formulation}
\label{subsec:continuum_description}
\noindent The continuum formulation of fluid flow in porous media is based on the approximation of a porous body by a collection of infinitely many infinitesimal points, $\mathbf{x}$, occupying a region $M \subset \R^3$ with boundary $\partial M$. The fluid volume density, $\rho(\mathbf{x})$, and the porosity (pore volume fraction), $\phi(\mathbf{x})$, are continuously distributed scalar fields, and the fluid velocity, $\mathbf{v}(\mathbf{x})$, is a continuously distributed vector field over $M$. The flow continuity at all $\mathbf{x} \in M$ is given by
\begin{equation}
    \label{eq:continuity_density}
    \frac{\partial (\rho\,\phi)}{\partial t} + \nabla \cdot (\rho\,\mathbf{v}) = 0,
\end{equation}
and the equation of fluid motion is given by
\begin{equation}
    \label{eq:motion_fluid}
    \mathbf{v} = - \frac{k}{\mu} \, \nabla p,
\end{equation}
where $p$ is the fluid pressure, $k$ is a pore space structure parameter referred to as the permeability, and $\mu$ is the fluid dynamic viscosity. \Cref{eq:motion_fluid} is the velocity version of the Darcy's law relating the volumetric flow rate to the gradient of pressure. The permeability is a tensorial quantity that can be a spherical tensor for isotropic media, or a general tensor for fully anisotropic media. We will refer to the ratio $K=k/\mu$ as the conductivity. It is related to the engineering hydraulic conductivity via $K_h=K\,g\,\rho$, where $g$ is the gravitational acceleration. The pressure is related to fluid density and porosity via equations of state for the fluid and the solid parts of the system, respectively, which contain material parameters -- compressibility of fluid, $C_F$, and of solid, $C_S$. For elastic transient flow the substitution of \cref{eq:motion_fluid} into \cref{eq:continuity_density} and considering the equations of state leads to (see, e.g., \cite{xue2021porous}):
\begin{equation}  \label{eq:continuous_governing_law}
  \frac{\partial (C\,p)}{\partial t} = \nabla \cdot (K\, \nabla p).
\end{equation}
where $C=\phi \, C_F + C_S$ is the total compressibility of the porous medium. The quantities of interest and their physical dimensions are summarised in \Cref{tab:continuous_quantities}. Note that $\mass$ denotes mass, $\length$ denotes length, and $\time$ denotes time.

\begin{table}[!ht]
  \caption{Quantities in the classical continuum formulation of flow in porous media}
  \label{tab:continuous_quantities}
  \centering
  \begin{tabular}{llll}
    \hline
    \noalign{\vskip .25em}
    Quantity & Symbol & Physical dimension & Field type \\
    \hline
    \noalign{\vskip .25em}
    fluid pressure & $p$ & $\mass \length^{-1} \time^{-2}$ & scalar \\
    fluid velocity & ${\bf v}$ & $\length \time^{-1}$ & vector \\
    fluid density & $\rho$ & $\mass \length^{-3}$ & scalar \\
    porosity & $\phi$ & $1$ & scalar \\
    permeability & $k$ & $\length^2$ & scalar/tensor \\
    dynamic viscosity & $\mu$ & $\mass \length^{-1} \time^{-1}$ & scalar \\
    conductivity & $K$ & $\mass^{-1} \length^3 \time$ & scalar/tensor \\
    compressibility of fluid/solid/total & $C_F/C_S/C$ & $\mass^{-1} \length \time^2$ & scalar \\
    \hline
  \end{tabular}
\end{table}

Initial-boundary value problems are constructed by prescription of initial and boundary conditions. The initial condition is a scalar function $p_0(\mathbf{x})$ prescribing the fluid pressure at time $t=0$ at all $\mathbf{x} \in M$, i.e., $p_0(\mathbf{x})=p(\mathbf{x}, t=0)$. The boundary $\partial M$ is partitioned into two non-overlapping parts $\Gamma_D$ and $\Gamma_N$, where Dirichlet and Neumann boundary conditions are prescribed by scalar functions $\Tilde{p}(\mathbf{x}, t)$ and $\Tilde{v}(\mathbf{x}, t)$, respectively, so that: $p(\mathbf{x}, t) = \Tilde{p}(\mathbf{x}, t)$, for $\mathbf{x} \in \Gamma_D$, and $\mathbf{v}(\mathbf{x}, t) \cdot \mathbf{n} = \Tilde{v}(\mathbf{x}, t)$, for $\mathbf{x} \in \Gamma_N$, where $\mathbf{n}$ is the outer unit normal to $\Gamma_N$.

In the following sub-sections we will describe the discrete analogues of all quantities and operations involved in the formulation of fluid flow and will arrive at a background-independent formulation of this process. 
Our approach is based on ideas from the smooth exterior calculus developed by \'Elie Cartan in 1899 \cite{cartan1899certaines}. Therefore, we will make a quick overview of this calculus, before introducing its discrete analogue and the discrete formulation.

\subsection{Overview of exterior calculus}
\label{subsec:exterior_calculus}
\noindent \textit{Exterior calculus} is an intrinsic (i.e., coordinate-independent) and metric-independent theory which forms the backbone of modern differential geometry.
It is a versatile theory that vastly generalises vector calculus by giving much more structured formalism for higher-dimensional calculus and geometry.
It is based on differential forms and its two main ingredients include the algebraic exterior product, also referred to as the wedge product, and the analytic exterior derivative.

\newterm{(Smooth) differential forms} are the objects of integration, see, e.g., \cite{bachman2012diffForms}, with many applications in differential geometry, see, e.g., \cite{fortney2018visualForms}.
As such they form a rigorous counterparts of the lousily defined infinitesimal line, area, and volume elements from multi-variable calculus.
A $p$-form on a manifold $M$ is represented as a linear combination of terms of the form
\begin{equation}
  f\, d x_{i_1} \wedge ... \wedge d x_{i_p},
\end{equation}
for some function $f$ on $M$ and some multi-index $I = (i_1, .., i_p)$.
The space of $p$-forms on $M$ is denoted by $\Omega^p M$.
On a smooth manifold of dimension $3$ they interpret:
\begin{itemize}
  \item a scalar field $f$ as a differential 0-form $f$;
  \item a vector field $(P, Q, R)$ as a differential 1-form $P\, d x + Q\, d y + R\, d z$;
  \item a pseudo-vector field $(P, Q, R)$ as a differential 2-form $P\, d y \wedge \, d z + Q\, d z \wedge \, d x + R\, d x \wedge \, d y$;
  \item a pseudo-scalar field $f$ as a differential 3-form $f\, dx \wedge d y \wedge \, d z$.
\end{itemize}

The \newterm{(smooth) exterior product} $\wedge$ of differential forms is an algebraic antisymmetric operation that takes a $p$-form $\omega$ and a $q$-form $\eta$ and returns a $(p + q)$-form $\omega \wedge \eta$, for non-negative integers $p$ and $q$ with $p + q \leq 3$.
The main algebraic operations in vector calculus are interpreted as follows:
\begin{itemize}
  \item multiplication of a vector field with a function is interpreted by the exterior product of a $0$-form with a $1$-form;
  \item dot product of vector fields is interpreted by the exterior product of a $1$-form with a $2$-form;
  \item cross product of vector fields is interpreted by the exterior product of a $1$-form with a $1$-form.
\end{itemize}
Note that $\eta \wedge \omega = (-1)^{p q} \omega \wedge \eta$ and so all exterior products have interpretations back in vector calculus.

The \newterm{(smooth) exterior derivative} $d$ of a differential form is an analytic operation that takes a $p$-form and returns a $(p + 1)$-form.
The main differential operators in vector calculus are interpreted as follows:
\begin{itemize}
  \item the gradient as the exterior derivative of a $0$-form;
  \item the curl as the exterior derivative of a $1$-form;
  \item the divergence as the exterior derivative of a $2$-form.
\end{itemize}
The operator $d$ satisfies the rule $d \circ d = 0$ and generalises the identities $\curl \circ \grad = 0$ and $\div \circ \curl = 0$.

\newterm{Integration of differential forms} assigns a real number to an oriented $p$-dimensional submanifold $S$ and a $p$-form $\omega$, denoted by $\int_S \omega$.
It generalises single, multiple, line and surface integrals.
If $\partial S$ denotes the oriented boundary of $S$, then the \textit{generalised Stokes' theorem} (or \textit{Stokes-Cartan theorem}) states that
\begin{equation}
  \int_{\partial S} \omega = \int_S d \omega
\end{equation}
for a $(p - 1)$-form $\omega$.
It generalises the following theorems:
\begin{itemize}
  \item Green's theorem for gradient ($p = 1$);
  \item Kelvin-Stokes theorem for curl ($p = 2$);
  \item Gauss's theorem for divergence ($p = 3$).
\end{itemize}

The operations discussed so far, exterior product, exterior derivative, and integration, do not depend on any notion of metric, although they interpret operations of vector calculus. A procedure for interchanging $p$-forms with $(3 - p)$-forms is needed to create an analogue of interchanging scalar with pseudo-scalar and vector with pseudo-vector fields in vector calculus. This requires the introduction of a \newterm{metric tensor}, fundamental in \textit{Riemannian geometry}, which is used in the construction of inner product of forms, Hodge star operators between forms and co-differentials.

The \newterm{inner product of differential forms} assigns a real number to $p$-forms $\omega$ and $\eta$, denoted by $\inner{\omega}{\eta}$.
The \newterm{Hodge star operator} takes a $p$-form $\omega$ and returns a $(3 - p)$-form $\star \omega$.
Hodge star operator satisfies $\star \star \omega = \omega$ and interprets the identification of scalar with pseudo-scalar fields by identifying $0$- with $3$-forms, and of vector with pseudo-vector fields by identifying $1$- with $2$-forms.
The Hodge-star and inner product are related by the equation
\begin{equation}
  \int_M (\omega \wedge \eta) = \inner{\omega}{\star \eta}
\end{equation}
for a $p$-form $\omega$ and a $(3 - p)$-form $\eta$.
In our case the Euclidean metric is chosen, which leads to the following equalities:
\begin{itemize}
  \item
    a $0$-form $\omega := f$ and a $3$-form $\eta := f\, dx \wedge d y \wedge \, d z$ are related by $\star_0 \omega = \eta,\ \star_3 \eta = \omega$;
  \item
    a $1$-form $\omega := P\, d x + Q\, d y + R\, d z$ and a $2$-form $\eta := P\, d y \wedge \, d z + Q\, d z \wedge \, d x + R\, d x \wedge \, d y$ are related by $\star_1 \omega = \eta,\ \star_2 \eta = \omega$;
\end{itemize}

The \newterm{codifferential} $d^\star$ takes a $p$-form and returns a $(p - 1)$-form.
Applied to a $(3 - p)$-form $\omega$, it satisfies the equation
\begin{equation}
  d^\star_p\, (\star_{3 - p}\, \omega) = (-1)^p \star_{4 - p}\, (d_{3 - p}\, \omega).
\end{equation}
By this definition it is easy to see that $d^\star_1$ can be used to define the analogue of divergence on a $1$-form.

\subsection{Exterior calculus formulation of fluid flow}
\label{subsec:exterior_governing_law}
\noindent Consider the right-hand side of \cref{eq:continuous_governing_law}.
The pressure $p$ can be represented as a $0$-form, and its gradient is substituted with the differential $d_0$ which leads to a $1$-form $d_0 p$.
The conductivity $K$ transforms $d_0 p$ to a $1$-form
\begin{equation}
  \label{eq:volumetric_flow_rate_2_form_dual}
  u := K\, d_0\, p,
\end{equation}
which is an analogue of the fluid velocity.
(More precisely, the fluid velocity ${\bf v}$ is the {\it sharp} of $u$, where sharp is the musical isomorphism from 1-forms to vector fields, induced by the inner product.)
As we previously said, the analogue of divergence on forms is $d_2$.
In order to apply $d_2$, the $1$-form $u$ has to be transformed into the \textit{volumetric flow rate} $2$-form $Q [\length^3 \time^{-1}]$,
\begin{equation}
  \label{eq:volumetric_flow_rate_2_form}
  Q := \star_1\, u.
\end{equation}
If $S$ is a oriented surface, then $\int_S Q$ represents the amount of volume flowing through $S$ at a unit of time. Finally, the resulting $3$-form $d_2 Q$ is transformed back into a $0$-form using $\star_3$. As a result, the analogue of divergence on $1$-forms, applied to $u^1$
is the operator $\star_3 \circ d_2 \circ \star_1 = - d_1^\star$.
This leads to the following exterior calculus analogue of \cref{eq:continuous_governing_law}:
\begin{equation}
  \label{eq:exterior_calculus_governing_law}
  \frac{\partial (C p)}{\partial t} = - d_1^\star\, K\, d_0\, p.
\end{equation}
In this formulation there is no change in the way the initial condition and the Dirichlet boundary condition are imposed.
The Neumann boundary condition, on the other hand, is imposed as
\begin{equation}
  \label{eq:neumann_bc_in_exterior_calculus}
  \tr_{\Gamma_N} Q = q(x, t),
\end{equation}
where $q(x, t) \in \Omega^2 \Gamma_N$ is the volumetric flow rate through the Neumann boundary $\Gamma_N$, $\tr_{\Gamma_N}$ is the restriction operator of a form on $M$ to a form on $\Gamma_N$ (more precisely, it is the pullback of the inclusion map from $\Gamma_N$ to $M$).
\Cref{tab:exterior_calculus_quantities} summarises the quantities participating in the governing equations in exterior calculus version.
Note that in this model the quantity that expresses the fluid flow is the physically meaningful volumetric flow rate $Q$ (albeit represented by its dual $u$), and not the fluid velocity ${\bf v}$.
Precisely, they are related by ${\bf v} = u^\sharp = (\star_2 Q)^\sharp$, where $\sharp$ is the \textit{musical isomorphism} that maps $1$-forms into vector fields.

\begin{table}[!ht]
  \caption{Quantities in the exterior calculus formulation of flow in porous media}
  \label{tab:exterior_calculus_quantities}
  \centering
  \begin{tabular}{llll}
    \hline
    \noalign{\vskip .25em}
    Quantity & Symbol & Physical dimension & Domain \\
    \hline
    \noalign{\vskip .25em}
    fluid pressure & $p$  & $\mass \length^{-1} \time^{-2}$ & $\Omega^0 M$ \\
    volumetric flow rate & $Q$ & $\length^3 \time^{-1}$ & $\Omega^2 M$ \\
    dual volumetric flow rate & $u$ & $\length^2 \time^{-1}$ & $\Omega^1 M$ \\
    conductivity & $K$ & $\mass^{-1} \length^3 \time$ & $\Omega^1 M \to \Omega^1 M$ \\
    compressibility & $C$ & $\mass^{-1} \length \time^2$ & $\Omega^0 M \to \Omega^0 M$ \\
    \hline
  \end{tabular}
\end{table}

In the following sub-sections we will present the discrete analogue of this exterior calculus formulation -- conservation law \cref{eq:exterior_calculus_governing_law}, supplied with initial, Dirichlet, and Neumann (\cref{eq:neumann_bc_in_exterior_calculus}) conditions.

\subsection{Discrete representation of porous and fractured media}
\label{subsec:discrete_description}
\noindent The smooth exterior calculus introduced briefly in \Cref{subsec:exterior_calculus} generalises the vector calculus and can replace it in modelling continuous media as shown in \Cref{subsec:exterior_governing_law}. However, as discussed in the introduction, porous and fractured media have complex microstructures, and our aim is to capture the complexity as closely as possible. For this we first need an appropriate mathematical description of microstructures and a discrete analogue of the smooth exterior calculus to represent the processes in such microstructures.

A material body with given internal structure can be described as a \newterm{cell complex} (or \newterm{mesh}) in the terminology of algebraic topology.
A cell complex $\mathcal{M}$ of dimension $3$, consists of $p$-cells, $0 \leq p \leq 3$, and let $N_p$ denote the number of $p$-cells in $\mathcal{M}$.
For $p = 0, 1, 2, 3$, the $p$-cells represent nodes (cells of dimension $0$), edges (cells of dimension $1$), surface patches (cells of dimension $2$), and bulk regions (cells of dimension $3$), respectively.
We will denote arbitrary $p$-cells by $a_p,\ b_p,\ c_p$.
The topological \newterm{boundary} of $c_p$ is a union of $0, 1, ..., (p-1)$-cells.
For $q < p$, if $b_q$ is on the boundary of $c_p$, we write $b_q \prec c_p$.
Several notions from algebraic topology are essential for our consideration.

A $p$-\newterm{chain} is a formal linear combination of $p$-cells in $\mathcal{M}$.
Thus, the space of all $p$-chains, denoted by $C_p \mathcal{M}$, has the set of of $p$-cells in $\mathcal{M}$ as a standard basis.
Hence, we will denote basis chains in $C_p \mathcal{M}$ by the same letters as for $p$-cells.
An arbitrary $p$-chain is a linear combination of basis $p$-chains, and we will use letters $\sigma_p,\ \tau_p$ for them.
The \newterm{relative orientations} assign $\pm 1$ to pairs $(c_p \succ b_{p - 1})$ of $p$-cells and their $(p - 1)$-dimensional boundary cells, denoted by $\varepsilon(c_p, b_{p - 1})$.
The assignment of relative orientations on $M$ is not unique (see Appendix B in the \href{https://ars.els-cdn.com/content/image/1-s2.0-S0307904X22002657-mmc1.pdf}{supplementary material} of \cite{berbatov2022diffusion} for a particular way of building it), it just needs to make sure that the two adjacent $0$-cells of a $1$-cell are opposite to one another, and makes the boundary operator $\partial$, defined shortly, satisfying $\partial^2 = 0$.
However, it is important to note that we assume that $\mathcal{M}$ is \newterm{compatibly oriented} in the sense that the relative orientations between any two adjacent volumes ($3$-cells) and their common $2$-cell are opposite to one another.
It is also helpful to assume that the boundary $\partial \mathcal{M}$ of $\mathcal{M}$ is compatibly oriented, i.e., adjacent $2$-cells on $\partial \mathcal{M}$ have opposite relative orientations with their common $1$-cell.

The \newterm{boundary operator} $\partial_p \colon C_p \mathcal{M} \to C_{p - 1} \mathcal{M}$ is a linear map from $p$-chains to $(p - 1)$-chains, given by its action on an arbitrary basis $p$-chain $c_p$ as
\begin{equation}
  \partial_p c_p := \sum_{b_{p - 1} \prec c_p} \varepsilon(c_p, b_{p - 1})\, b_{p - 1},
\end{equation}
and extended by linearity on all $p$-chains.
In the standard basis of basis chains, $\partial_p$ is a sparse matrix with dimension $N_{p-1} \times N_p$ and nonzero values $\pm 1$ representing relative orientations.
The boundary operator satisfies the equation $\partial_p \, \circ \, \partial_{p + 1} = 0$, i.e., the boundary of a boundary is empty, which in algebraic terminology means that the collection of spaces of $p$-chains, denoted by $C_\bullet  \mathcal{M}$, forms, together with the boundary operator, a \newterm{chain complex} $(C_\bullet  \mathcal{M}, \partial)$.

A linear function on $p$-chains is called a $p$-\newterm{cochain}.
A $p$-cochain can be regarded as a function assigning a real number to each $p$-cell in $\mathcal{M}$.
The $p$-cochains in $\mathcal{M}$ form a vector space denoted by $C^p \mathcal{M}$ (the \newterm{linear-algebraic dual space} of $C^p \mathcal{M}$).
In coordinates, a $p$-cochain is a column vector with dimension $N_p$.
A basis $p$-cochain, corresponding to a cell $c_p$, is denoted by $c^p$, and is $1$ when applied to $c_p$, and zero elsewhere.
General $p$-cochain will be denoted by $\sigma^p,\ \tau^p$.

The \newterm{linear-algebraic dual map} of $\partial_{p + 1}$ is the \newterm{coboundary operator} $\delta_p \colon C^p  \mathcal{M} \to C^{p + 1}  \mathcal{M}$, which maps $p$-cochains to $(p + 1)$-cochains. Formally,
\begin{equation}
  \delta_p \sigma^p := \sigma^p \circ \partial_{p + 1}.
\end{equation}
In the standard basis, $\delta_p$ is represented as a sparse matrix of dimension $N_{p+1} \times N_{p}$ and nonzero values $\pm 1$; it is the transpose matrix of $\partial_{p + 1}$.
The coboundary operator satisfies the equation $\delta_{p + 1} \, \circ \, \delta_p = 0$, i.e., the coboundary of coboundary is empty.
This makes the collection of spaces of $p$-cochains, denoted by $C^\bullet  \mathcal{M}$, together with the coboundary operator, a \newterm{cochain complex} $(C^\bullet  \mathcal{M}, \delta)$.

The notions discussed so far are prevalent in algebraic topology and correspond to smooth exterior calculus in the following way:
\begin{itemize}
  \item a cell complex $\mathcal{M}$ is a finite version of the collection of all sub-manifolds of the ambient manifold $M$;
  \item a chain in $\mathcal{M}$ corresponds to a formal sum of oriented sub-manifolds of $M$;
  \item the boundary operator $\partial$ on $\mathcal{M}$ corresponds to the oriented boundary on chains in $M$;
  \item cochains on $\mathcal{M}$ correspond  to smooth differential forms on $M$.
    In the ``discretisation direction'', the \newterm{de Rham map} $ R \colon \Omega^p M \to C^p \mathcal{M}$ is defined by
    \begin{equation}
      (R\, \omega^p)(c_p) := \int_{c_p} \omega^p,
    \end{equation}
    where we interpret the integral over a basis chain $c_p$ as integral over the ambient sub-manifold it occupies.
    In the opposite ``interpolation direction'' cochains are mapped to piecewise-smooth forms by the so-called \newterm{Whitney forms} (see, e.g., \cite[Section 4]{wilson2007cochain} for simplicial and \cite[Section 3.2.2]{arnold2012discrete} for cubical Whitney forms);
  \item the coboundary operator $\delta$ on $\mathcal{M}$ corresponds to the smooth exterior derivative $d$ on $M$.
    They are related by the equality
    \begin{equation}
      R \circ d = \delta \circ R.
    \end{equation}
\end{itemize}

A discrete analogue of the smoooth wedge product, if it exists, is called the \newterm{cup product} of cochains. Cup products have been proposed for simplicial \cite[Definition 5.1]{wilson2007cochain}, cubical \cite[Definition 3.2.1]{arnold2012discrete} and polygonal \cite[Definition 3.2.4]{ptackova2017discrete} meshes. However, we are not aware of discrete analogues of the wedge product for more general types of meshes, typically needed for faithful representation of materials' internal structures. Therefore, we proceed in a different direction using Forman's combinatorial differential forms \cite{forman2002combinatorial} and Forman's isomorphism between these forms and the cochains of an extended complex $\mathcal{K}$. This isomorphism allows to pull back a cup product of cochains on $\mathcal{K}$, if it exists, to a discrete wedge product of forms on $\mathcal{M}$. A larger class of $\mathcal{M}$ topologies produce quasi-cubical $\mathcal{K}$, hence allow for a well defined cup product of cochains on $\mathcal{K}$, and thus a discrete wedge product of combinatorial forms on $\mathcal{M}$. Importantly, the combinatorial forms allow for more complex interactions between cells, which is essential for our purposes.

\begin{figure}[!ht]
  \begin{subfigure}{.5\textwidth}
    \centering
    \includegraphics[scale=.2]{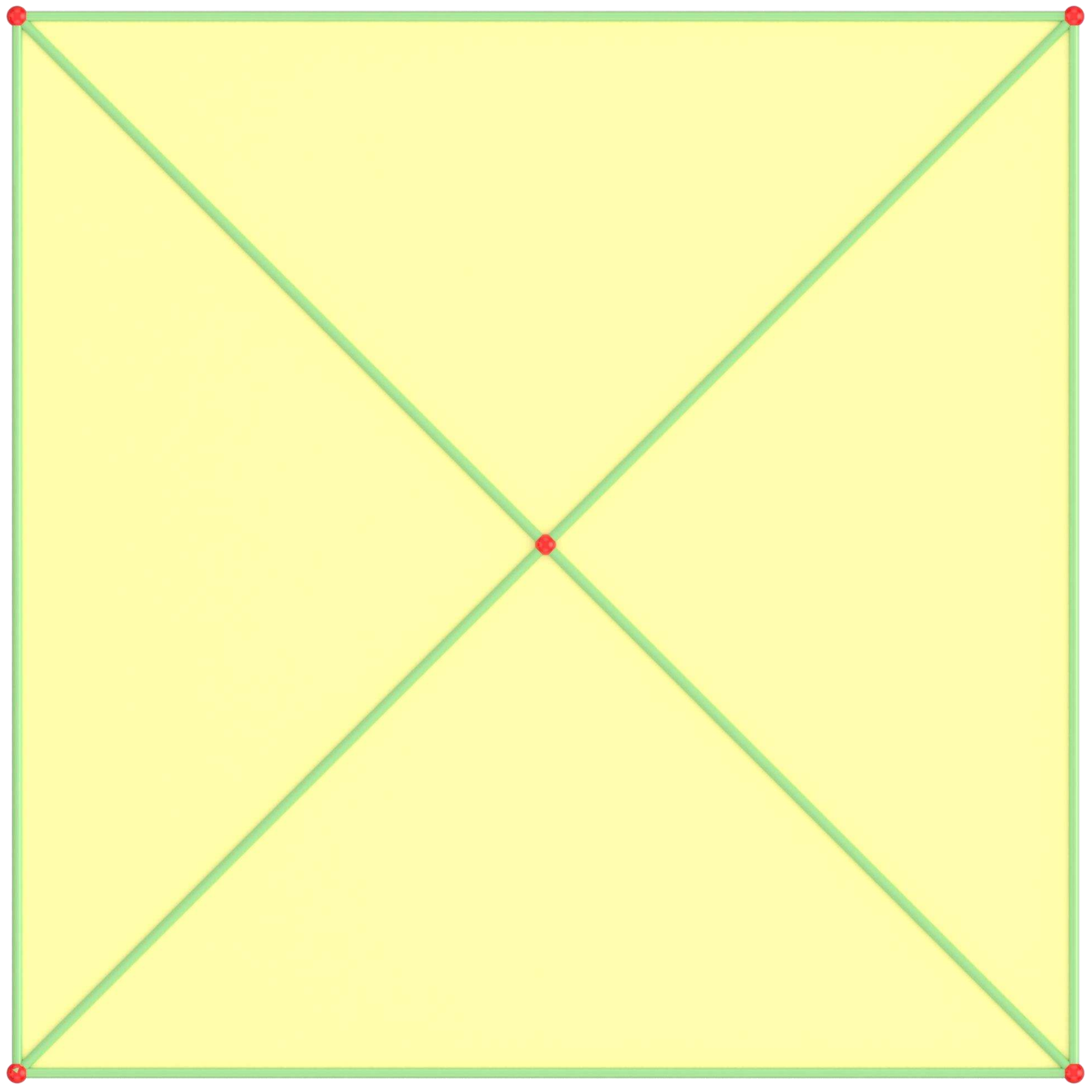}
    \caption{}
  \end{subfigure} 
  \begin{subfigure}{.5\textwidth}
    \centering
    \includegraphics[scale=.2]{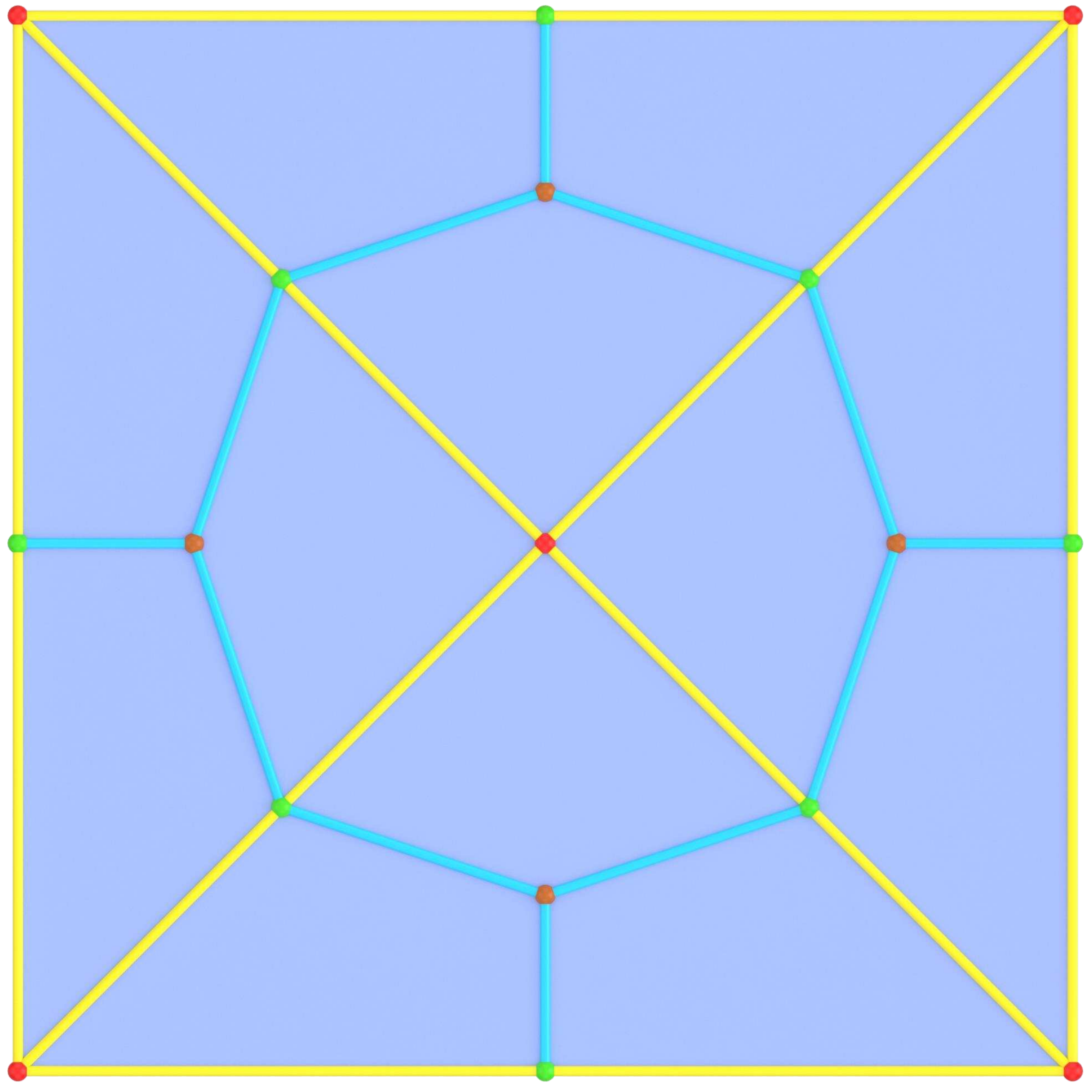}
    \caption{}
  \end{subfigure}
  \caption{Triangular mesh $\mathcal{M}$ (a) and its Forman subdivision $\mathcal{K}$ (b)}
  \label{fig:triangulation_0p3}
\end{figure}

A \newterm{combinatorial differential $q$-form}, denoted by $\omega^q$, is a mapping between cells. For $p \geq q$, $\omega^q$ maps a $p$-cell $c_p$ to a linear combination of the $(p - q)$-cells on the boundary of $c_p$. Thus, $\omega^q$ can be interpreted as an assignment of numbers to all pairs $(b_{p - q} \preceq c_p)$.
The space of all $q$-forms will be denoted by $\Omega^q \mathcal{M}$.
The \newterm{discrete exterior derivative} $D_q \colon \Omega^q \mathcal{M} \to \Omega^{q + 1} \mathcal{M}$ is defined by \cite[Definition 2.6]{berbatov2022diffusion}
\begin{equation}
  D_q \omega^q := \omega^q \circ \partial - (-1)^p \partial \circ \omega.
\end{equation}
(This definition differs by sign to the original \cite[Definition 1.1]{forman2002combinatorial}.)
It is easy to show \cite[Proposition 2.8]{berbatov2022diffusion} that $D_{p + 1} \circ D_p = 0$ which makes the space of all combinatorial differential forms, together with the discrete exterior derivative, a cochain complex, denoted by $(\Omega^\bullet \mathcal{M}, D)$.
Forman \cite[Theorem 1.2]{forman2002combinatorial} proved that the \newterm{cohomology} of $(\Omega^\bullet \mathcal{M}, D)$ is isomorphic to the cohomology of $(C^\bullet \mathcal{M}, \delta)$, where $\delta$ is the coboundary operator on cochains in $\mathcal{M}$.
He gave an algebraic proof but sketched a geometric proof by building a subdivision $\mathcal{K}$ of $\mathcal{M}$ such that $(\Omega^\bullet \mathcal{M}, D) \cong (C^\bullet \mathcal{K}, \delta)$, where $\delta$ is the coboundary operator on cochains in $\mathcal{K}$. In such way, the combinatorial forms in $\mathcal{M}$ are interpreted as cochains in $\mathcal{K}$, and the exterior derivative of forms in $\mathcal{M}$ is given by the the coboundary operator in $\mathcal{K}$.
This subdivision, referred to as the \newterm{Forman subdivision}, was later studied in \cite[Chapter 5]{arnold2012discrete}, referred to as the \newterm{kite complex}, again for topological purposes.
Recently, the Forman subdivision was shown to be a suitable foundation for calculus on meshes and consequently for modelling materials with complex internal structures \cite[Section 4]{berbatov2022diffusion}. The present work demonstrates its computational capabilities in modelling fluid flow in porous media.
The Forman subdivision is built as follows:
\begin{itemize}
  \item
    a $p$-cell in $\mathcal{K}$ corresponds to a pair of a $q$-cell $b_q$ in $\mathcal{M}$ ($q \geq p$) and a $(q - p)$-cell $a_{q - p}$ on the boundary of $b_q$; this cell is denoted by $(b_q \to a_{q - p})$
  \item
    the topology on $\mathcal{K}$, represented by the boundary relation $\prec$, is defined by considering the following cells in $\mathcal{K}$ for $q > p$: a $p$-cell $\alpha_p := (c_r \to b_{r - p})$ and a $q$-cell $\beta_q := (d_s \to a_{s - q})$; then, by definition, $\alpha_p \prec \beta_q$ whenever $a_{s - q} \preceq b_{r - p}$ and $c_r \preceq d_s$;
  \item
    relative orientations in $\mathcal{K}$ are defined by relative orientations in $\mathcal{M}$, see \cite[Equation (2.3)]{berbatov2022diffusion};
  \item
    coordinates of nodes in $\mathcal{K}$ can be defined in many ways: we use centroids of cells in $\mathcal{M}$.
\end{itemize}
The Forman subdivision is topologically unique. It is illustrated in \Cref{fig:triangulation_0p3} on a small $2D$ triangular mesh.
Note, that different colours in the subdivision correspond to pairs of different types in the original mesh.
In $\mathcal{M}$ nodes are in red, edges are in green, faces are in yellow.
Correspondingly, nodes in $\mathcal{K}$ are in red/green/orange respectively, depending on the type of cell they correspond to in $\mathcal{M}$.
Edges in $\mathcal{K}$ are in blue (corresponding to pairs $b_1 \succ a_0$) and in yellow (corresponding to pairs $b_2 \succ a_1$).
Faces in $\mathcal{K}$ are in pink, corresponding to pairs $b_2 \succ a_0$.

In the 3D case, the construction leads to three types of 1-cells in $\mathcal{K}$: from 3-cells to 2-cells in $\mathcal{M}$ corresponding to pairs $b_2 \prec c_3$; from 2-cells to 1-cells in $\mathcal{M}$ corresponding to pairs $b_1 \prec c_2$; and from 1-cells to 0-cells in $\mathcal{M}$ corresponding to pairs $b_0 \prec c_1$. If $\mathcal{M}$ is a model of a material domain containing components of different apparent dimensions, the three types of 1-cells in $\mathcal{K}$ allow for associating different properties to different dimensional components and for processes to operate simultaneously on all components. This means that voids and grains can be allocated to different dimensional cells according to their geometries, a marked computational advantage over the continuum theory. Furthermore, there are two types of 2-cells in $\mathcal{K}$: from 3-cells to 1-cells in $\mathcal{M}$ representing pairs $b_1 \prec c_3$; and from 2-cells to 0-cells in $\mathcal{M}$ representing pairs $b_0 \prec c_2$.

\subsection{Calculus on quasi-cubical cell complexes}
\label{subsec:discrete_calculus}
\noindent In the $3$D case the extended complex $\mathcal{K}$ consists of topological cubes (\newterm{quasi-cubes}) when $\mathcal{M}$ is formed of simple polyhedrons, i.e., polyhedrons with $3$ edges coming out of every vertex. The class of simple polyhedrons is large, including tetrahedrons, hexahedrons, and polyhedrons arising from space tessellations, such as Voronoi.
On the quasi-cubes we apply the cup product $\smile$ of \cite[Definition 2.14]{berbatov2022diffusion}, generalised from \cite[Definition 3.2.1]{arnold2012discrete}.
Since forms on $\mathcal{M}$ are equivalent to cochains in $\mathcal{K}$, the cup product on $\mathcal{K}$ is interpreted as a \newterm{discrete wedge product} on $\mathcal{M}$.

Two cells $b_{3 - p}, c_p \in \mathcal{K}$ are \newterm{topologically orthogonal} if they share only one common $0$-cell and are boundaries of a common $3$-cells. This is written as $c_p \perp b_q$.
The \newterm{discrete inner product} on $\mathcal{K}$ is a bilinear form on $C^p \mathcal{K}$ such that basis cochains form an orthogonal basis, and for a $p$-cell $c_p$, \cite[Equation (2.90)]{berbatov2023discrete}
\begin{equation}
  \inner{c^p}{c^p} := \frac{1}{8 \mu(c_p)} \sum_{b_{3 - p} \perp c_p} \mu(b_{3 - p}),
\end{equation}
where $\mu(a_q)$ denotes the measure of the $q$-cell $a_q$ (always $1$ for $q = 0$, length for $q = 1$, area for $q = 2$, and volume for $q = 3$).

The \newterm{discrete Hodge star operator}, denoted by $\star_p$, is the unique operator satisfying \cite[Definition 3.3]{berbatov2022diffusion}
\begin{equation}
  (\tau^{3 - p} \smile \sigma^p)[\mathcal{K}] = \inner{\tau^{3 - p}}{\star_p \, \sigma^p},
\end{equation}
for arbitrary cochains $\sigma^p \in C^p \mathcal{K}$ and $\tau^{3 - p} \in C^{3 - p} \mathcal{K}$. Here, $[\mathcal{K}]$ is the fundamental class of $\mathcal{K}$, i.e., the sum of all basis $3$-cochains.
Note, that the left hand side expresses a discrete form of integration.
The discrete Hodge star $\star_p$ is represented as a sparse matrix with dimension $N_{d-p} \times N_p$.

The \newterm{adjoint coboundary operator} $\delta^\star_p$ is the discrete analogue of the codifferential.
It is defined as the adjoint of $\delta_{p - 1}$ with respect to the inner product \cite[Definition 3.1]{berbatov2022diffusion}, that is
\begin{equation}
  \inner{\sigma^p}{\delta_{p - 1} \tau^{p - 1}} = \inner{\delta_p^\star \sigma^p}{\tau^{p - 1}},
\end{equation}
for arbitrary cochains $\sigma^p \in C^p \mathcal{K}$ and $\tau^{p - 1} \in C^{p - 1} \mathcal{K}$.
The adjoint coboundary operator $\delta^\star_p$ is represented as a sparse matrix with dimension $N_{p-1} \times N_p$ with the same stencil as the boundary operator $\partial_p$ (only magnitudes of values differ from those of the boundary operator; the signs are the same).

\begin{figure}[!ht]
  \centering
  \includegraphics[width=0.45\linewidth]{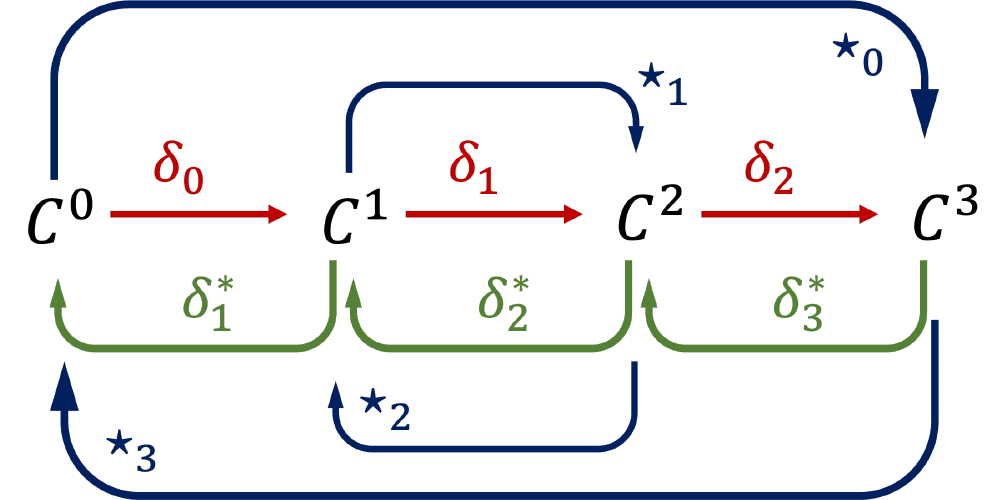}
  \caption{The cochain complex $(C^\bullet \mathcal{K}, \delta)$ and the metric-dependent operations on the cochain spaces.}
  \label{fig:CochainComplex}
\end{figure}

The operations introduced above for the cochain spaces in $\mathcal{K}$  are illustrated in \Cref{fig:CochainComplex}. Similarly to the interpretation of $\delta_0$, $\delta_1$, and $\delta_2$ as gradient, curl, and divergence, of scalar, vector, and pseudo-vector fields, respectively, the adjoint coboundary operators $\delta^\star_3$, $\delta^\star_2$, and $\delta^\star_1$ are interpreted as gradient, curl, and divergence, of pseudo-scalar, pseudo-vector, and vector fields, respectively. The relations between scalar and pseudo-scalar fields, and between vector and pseudo-vector fields are given by the Hodge-star operators. The algebraic representations of all discrete fields, i.e., all cochains, are 1-dimensional arrays, and the algebraic representations of all operators are 2-dimensional arrays (sparse matrices).

\subsection{Fluid flow in discrete media}
\label{subsec:discrete_flow}
\noindent The apparatus developed in \Cref{subsec:discrete_calculus} can be used for describing the problem of fluid flow based on the strong formulation given in \Cref{subsec:exterior_governing_law} or on a weak (integral) formulation. Here we demonstrate the discrete version of the strong formulation. 

The pressure $p$ is represented by a $0$-form on $\mathcal{M}$, i.e., pressure is defined for all cells in $\mathcal{M}$, hence it is a $0$-cochain on $\mathcal{K}$. Let this cochain be denoted by $\pi^0$. Further, the compressibility $\mathrm{C}$ is a material property that can be individually assigned to all cells in $\mathcal{M}$, so that $\mathrm{C}=\mathrm{C_F}$ or $\mathrm{C}=\mathrm{C_S}$ depending on whether the cell is associated with a void or a solid (see \cref{eq:continuous_governing_law}). This is transported as properties of $0$-cells in $\mathcal{K}$, forming a diagonal matrix $\Pi_0$ with dimension $N_0 \times N_0$ of prescribed local compressibilities.

The exterior derivative of the pressure $0$-form on $\mathcal{M}$ gives the pressure differentials between pairs of cells $(d_0 \prec c_1)$, $(c_1 \prec b_2)$, and $(b_2 \prec a_3)$ on $\mathcal{M}$. Correspondingly, $\delta_0 \, \pi^0$ gives a $1$-cochain of pressure differentials on $\mathcal{K}$. These differentials drive the flow between the cells of the corresponding pairs, that is along $1$-, $2$-, or $3$-cells in $\mathcal{M}$, with conductivities that can be different for different pairs depending on the local geometry. The local conductivities prescribed for transport between pairs in $\mathcal{M}$ are transported as local conductivities of the $1$-cells in $\mathcal{K}$, forming a diagonal matrix $\Pi_1$ with dimension $N_1 \times N_1$ of prescribed local conductivities.
Define the $1$-cochain $\upsilon^1$ on $\mathcal{K}$ by (compare with \cref{eq:volumetric_flow_rate_2_form_dual})
\begin{equation}
\label{eq:discrete_motion}
  \upsilon^1 := \Pi_1 \, \delta_0 \pi^0.
\end{equation}
$\upsilon^1$ does not relate to fluid velocity in a direct way, but like in the exterior calculus case its Hodge dual, $\psi^2 := \star_1 \, \upsilon^1$, is the discrete volumetric flow rate. 
More precisely, $\psi^2$ measures the fluid volumes passing the $2$-cells per unit of time.

The discrete continuity equation is then represented in an analogous to \cref{eq:exterior_calculus_governing_law} way:
\begin{equation}
\label{eq:discrete_balance}
  \frac{\partial (\Pi_0 \, \pi^0)}{\partial t}
  = -\delta^\star_1 \, \upsilon^1
  = -\delta^\star_1 \, \Pi_1 \, \delta_0 \pi^0.
\end{equation}
The action of $\delta^\star_1$ restores the balance of spatial dimensions to provide the discrete analogue of $\Delta \cdot \mathbf{v}$.
\Cref{eq:discrete_balance} can also be written with a material-modified Laplacian $\Tilde{\Delta}_0 := \delta^\star_1 \, \Pi_1 \, \delta_0$, which is a sparse matrix with dimension $N_0 \times N_0$:
\begin{equation}
\label{eq:discrete_Laplace}
  \frac{\partial (\Pi_0 \, \pi^0)}{\partial t} = - \Tilde{\Delta}_0 \pi^0.
\end{equation}
\Cref{tab:discrete_quantities} summarises the quantities participating in the governing equations of our discrete model (compare with \Cref{tab:exterior_calculus_quantities}).

\begin{table}[!ht]
  \caption{Quantities in the discrete formulation of flow in porous media}
  \label{tab:discrete_quantities}
  \centering
  \begin{tabular}{llll}
    \hline
    \noalign{\vskip .25em}
    Quantity & Symbol & Physical dimension & Domain \\
    \hline
    \noalign{\vskip .25em}
    fluid pressure & $\pi^0$  & $\mass \length^{-1} \time^{-2}$ & $C^0 \mathcal{K}$ \\
    volumetric flow rate & $\psi^2$ & $\length^3 \time^{-1}$ & $C^2 \mathcal{K}$ \\
    dual volumetric flow rate & $\upsilon^1$ & $\length^2 \time^{-1}$ & $C^1 \mathcal{K}$ \\
    conductivity (diagonal matrix) & $\Pi_1$ & $\mass^{-1} \length^3 \time$ & $C^1 \mathcal{K} \to C^1 \mathcal{K}$ \\
    compressibility (diagonal matrix) & $\Pi_0$ & $\mass^{-1} \length \time^2$ & $C^0 \mathcal{K} \to C^0 \mathcal{K}$ \\
    \hline
  \end{tabular}
\end{table}

Initial-boundary value problems for \cref{eq:discrete_balance} on a material cell complex $\mathcal{M}$ are prescribed as follows.
The boundary $\partial \mathcal{M}$ contains cells of $\mathcal{M}$ with dimensions less than $3$, the boundary $2$-cells and the edges and nodes they contain, and it itself is a $2$-dimensional cell complex.
Partition the set $(\partial \mathcal{M})_2$ of $2$-cells of $\partial \mathcal{M}$ into (disjoint) Dirichlet and Neumann parts, respectively $\Sigma_D^2$ and $\Sigma_N^2$.
Take all lower-dimensional cells they contain to arrive to the $2$-dimensional cell complexes $\Sigma_D$ and $\Sigma_N$.
Note that usually $\Sigma_D$ and $\Sigma_N$ will share common edges and nodes, where respective $2$-cells intersect, unless they are on different connected components.
The partitioning is transported to the boundary of $\mathcal{K}$, forming the sets $\Gamma_D$ and $\Gamma_N$, such that for $p = 0, 1, 2$,
\begin{equation}
\label{eq:Dirichlet_p_K}
  \Gamma_D^p := \{\alpha_p := c_q \to b_{q - p} \mid q \geq p,\ c_q \in \Sigma_D^q,\ b_{q - p} \preceq c_q\},
\end{equation}
\begin{equation}
\label{eq:Neumann_p_K}
  \Gamma_N^p := \{\alpha_p := c_q \to b_{q - p} \mid q \geq p,\ c_q \in \Sigma_N^q,\ b_{q - p} \preceq c_q\}.
\end{equation}
The conditions are prescribed as:
\begin{itemize}
  \item
    Initial: for initial pressure $p_0$ on $\mathcal{K}$,
    \begin{equation}
      \label{eq:Initial}
      \pi^0(c_0, 0)= p_0(c_0), \; c_0 \in \mathcal{K}_0 \cong \mathcal{M}.
    \end{equation}
  \item
    Dirichlet: for prescribed pressure $p$ on $\Gamma_D$,
    \begin{equation}
      \label{eq:Dirichlet}
      \pi^0(c_0, t)= p(c_0, t), \; c_0 \in \Gamma_D^0,\ t \geq 0.
    \end{equation}
  \item
    Neumann (compare with \cref{eq:neumann_bc_in_exterior_calculus}): for prescribed volumetric flow rate $f$ on $\Gamma_N$,
    \begin{equation}
      \label{eq:Neumann}
      \psi^2(c_2, t)= f(c_2, t), \; c_2 \in \Gamma_N^2,\ t \geq 0.
    \end{equation}
\end{itemize}

In our formulation, the number of unknowns at each time step is $N_0$ -- the number of coefficient representing the pressure $0$-cochain $\pi^0 \in C^0 \mathcal{K}$.
The equations available at each time step are: the conservation law for all internal nodes, Dirichlet condition for Dirichlet boundary, and Neumann condition for Neumann boundary. To achieve equal number of equations and unknowns, \cref{eq:Neumann} is modified for application on $0$-cells, using the Hodge star operator
\begin{equation}
  \star_{\Gamma_N, 2} \colon C^2 \Gamma_N \to C^0 \Gamma_N,
\end{equation}
which follows the generic construction from \cite{berbatov2023discrete}.
Taking only the $0$-cells in $\Gamma_N$ that are not in $\Gamma_D$ (Dirichlet condition is applied on common nodes), the Neumann boundary condition reduces to
\begin{equation}
  (\star_{\Gamma_N, 2}\, \tr_{\Gamma_N, 2}\, \star_1\, \Pi_1\, \delta_0\, \pi^0)(c_0, t) = (\star_{\Gamma_N, 2}\, f)(c_0, t),\ c_0 \in \Gamma_N^0 \setminus \Gamma_D^0,
\end{equation}
where $\tr_{\Gamma_N, 2}$ is the restriction of a $2$-form in $\mathcal{K}$ to a $2$-form on $\Gamma_N$.
To summarise, denoting by $\mathcal{K}^{\circ}_0$ the set of internal nodes of $\mathcal{K}$, we get the the following problem for the unknown time-dependent $0$-cochain $\pi^0$:
\begin{subequations}
  \label{eq:discrete_formulation}
  \begin{alignat}{3}
    & \frac{\partial \Pi_0 \pi^0}{\partial t}(c_0, t)
    && = - (\delta^\star_1 \circ \Pi_1 \circ \delta_0)(\pi^0)(c_0, t),\ \forall c_0 \in \mathcal{K}^{\circ}_0,\ t > 0 \qquad
    && [\time^{-1}], \\
    & \pi^0(c_0, t)
    && = p(c_0, t),\ \forall c_0 \in \Gamma_D^0,\ t \geq 0 \qquad
    && [\mass \length^{-1} \time^{-2}], \\
    & (\star_{\Gamma_N, 2} \circ \tr_{\Gamma_N, 2} \circ \star_1
      \circ \pi_1 \circ \delta_0)(\pi^0)(c_0, t)
    && = (\star_{\Gamma_N, 2}\, f)(c_0, t),\ \forall c_0 \in \Gamma_N^0 \setminus \Gamma_D^0,\ t \geq 0 \qquad
    && [\length \time^{-1}], \\
    & \pi^0(c_0, 0)
    && = p_0(c_0),\ \forall c_0 \in \mathcal{K}_0 \qquad
    && [\mass \length^{-1} \time^{-2}].
  \end{alignat}
\end{subequations}
Solving \cref{eq:discrete_formulation} requires a temporal numerical scheme, and then at each time leads to a linear system for the coefficients of $\pi^0$.
Imposing Dirichlet boundary conditions leads to exact numerical values for the coefficients of $\pi^0$ on the Dirichlet boundary.
The remaining system, however, exhibits equations in different physical dimensions (namely, $\time^{-1}$ for conservation law and $\length \time^{-1}$ for Neumann boundary condition), which is undesirable when forming the final left-hand side matrix and right-hand side vector.
For this reason, we suggest for every $c_0 \in \Gamma_N^0 \setminus \Gamma_D^0$ to divide the corresponding row in the final system by the arithmetic mean of the adjacent edges of $c_0$ in $\Gamma_N^1$.
This will lead to a linear system $A x = b$ with $N_0 - |\Gamma_D^0|$ unknowns and equations, where $A [\mass^{-1} \length \time],\ x[\mass \length^{-1} \time^{-2}],\ b[\time^{-1}]$.

Hereafter, the proposed approach will be used for calculating the conductivity and permeability of models of porous and fractured media. For this purpose we mimic experimental setups with specimens of simple geometry (cuboids or cylinders) where pressures $p_1 < p_2$ are applied on two opposite parallel sides of equal areas $A$ and at distance $L$, and the volumetric flow rate is measured. Typically, in such experiments the remaining sides of the specimens are insulated, i.e., the volumetric flow rate is zero. Solving the corresponding problem on a cell complex $\mathcal{M}$ provides the volumetric flow rates across the $2$-cells on the two opposite parallel boundaries with applied pressures. The conductivity is then calculated as in experiments by: 
\begin{equation}
  K = \frac{Q \cdot L}{(p_2 - p_1) \cdot A},
\end{equation}
where $Q$ is the sum of volumetric flow rates across one boundary. 

The solution to transient problems is performed by the implicit Euler method. The transient solutions converge to steady-state solutions in all cases, only the convergent times depend on the topological structure, geometry and material properties of $\mathcal{M}$. Since we are interested in calculating the conductivity, the results are obtained directly with steady-state solutions for computational efficiency. At steady-state, the sums of the volumetric flow rates at the two opposite sides are equal.

\section{Materials and models}
\label{sec:materials_and_models}
\noindent The application case studies presented in this section deal with fluid flow in a rock. Four different rocks are considered for demonstration of the proposed method: Bentheimer sandstone (BS), Doddington sandstone (DS), Estaillades carbonate (EC), and Ketton carbonate (KC). These are typical geomaterials with random and complex pore system and grain size distribution, making them suitable for demonstrating the capabilities of the new method. Bulk properties and XCT images of the four rocks are obtained from the dataset of the Department of Earth Science and Engineering, Imperial College London: \url{https://www.imperial.ac.uk/earth-science/} \cite{raeini2017generalized,andrew2014reservoir}. The bulk porosity and permeability of all rocks are given in \Cref{tab:rocks_experimental}. It can be seen from the data that the proportions between permeability and porosity vary significantly between different rocks, demonstrating that the porosity alone is not controlling the permeability. One needs to further consider the effects of pore shape and volume distribution for predicting permeability.

\begin{table}[!ht]
  \caption{Measured characteristics of rocks by bulk experiments}
  \label{tab:rocks_experimental}
  \centering
  \begin{tabular}{ccc}
    \hline
    \noalign{\vskip .25em}
    Material & Measured Helium porosity & Measured permeability ($10^{-12}\, m^2$) \\
    \hline
    \noalign{\vskip .25em}
    Bentheimer sandstone & 0.20 & 1.875 \\[.25em]
    Doddington sandstone & 0.192 & 1.038 \\[.25em]
    Estaillades carbonate & 0.295 & 0.149 \\[.25em]
    Ketton carbonate & 0.2337 & 2.807 \\[.25em]
    \hline
  \end{tabular}
\end{table}

\subsection{Material fabric's characteristics}
\begin{figure}[!ht]
\centering
    \begin{subfigure}{.20\textwidth}  
        \centering
        \includegraphics[width=\textwidth]{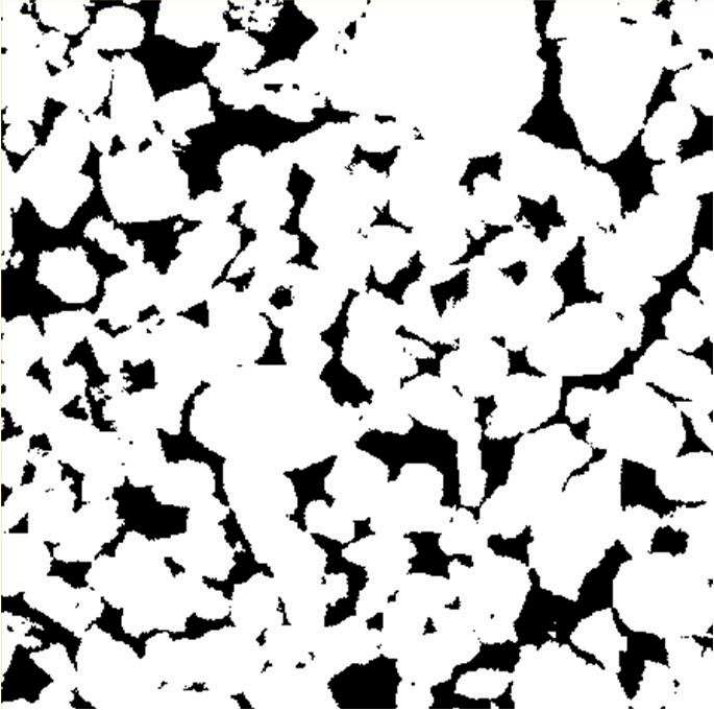}
        \caption{}
    \end{subfigure}
    \hspace{0.2em}  
    \begin{subfigure}{.20\textwidth}
        \centering
        \includegraphics[width=\textwidth]{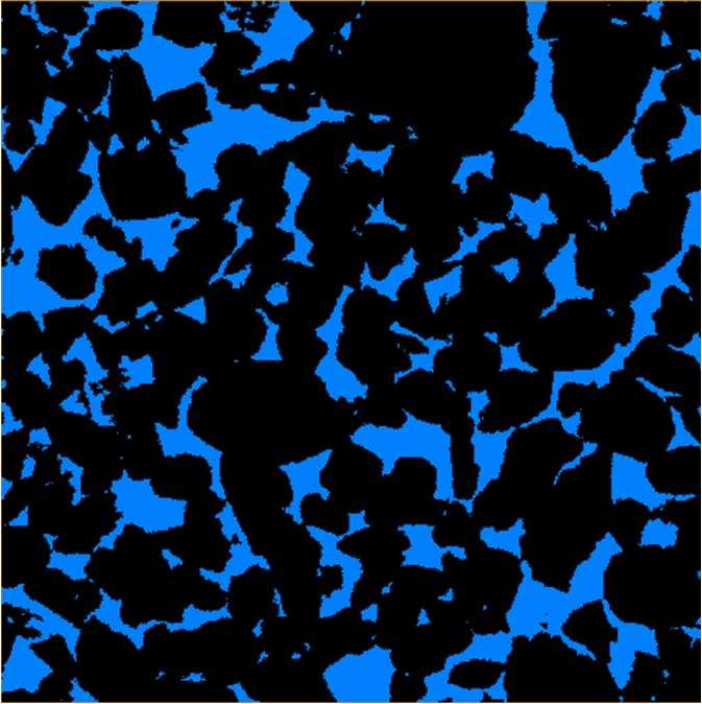}
        \caption{}
    \end{subfigure}
    
    \begin{subfigure}{.20\textwidth}
        \centering
        \includegraphics[width=\textwidth]{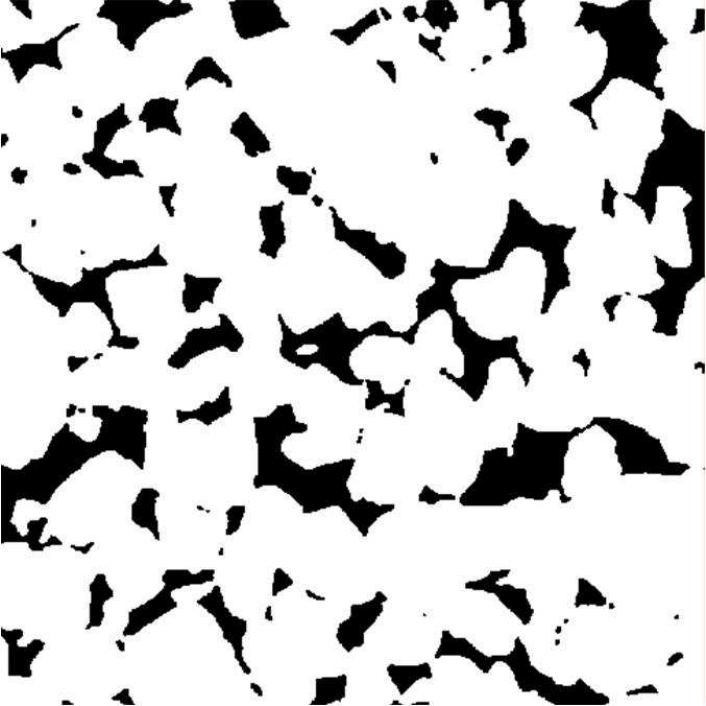}
        \caption{}
    \end{subfigure}
    \hspace{0.2em}
    \begin{subfigure}{.20\textwidth}
        \centering
        \includegraphics[width=\textwidth]{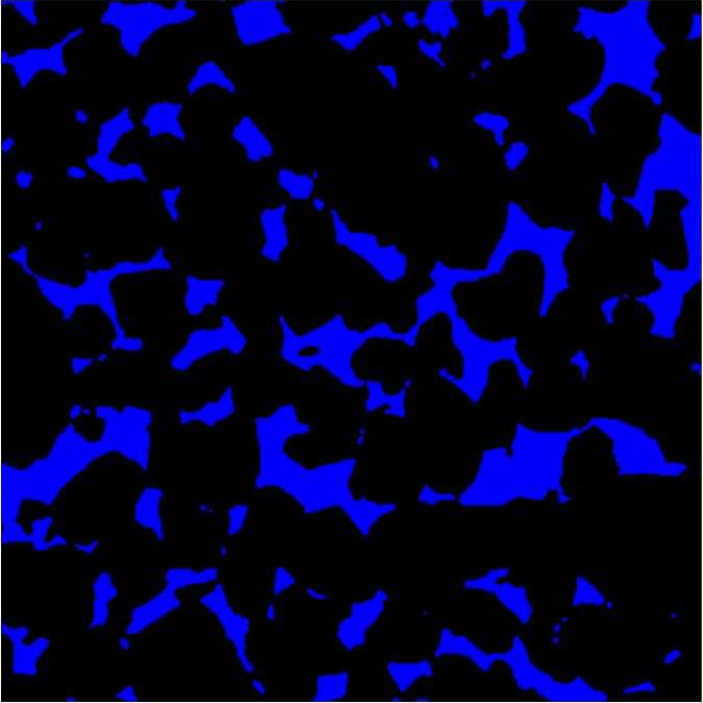}
        \caption{}
    \end{subfigure}

    \begin{subfigure}{.20\textwidth}
        \centering
        \includegraphics[width=\textwidth]{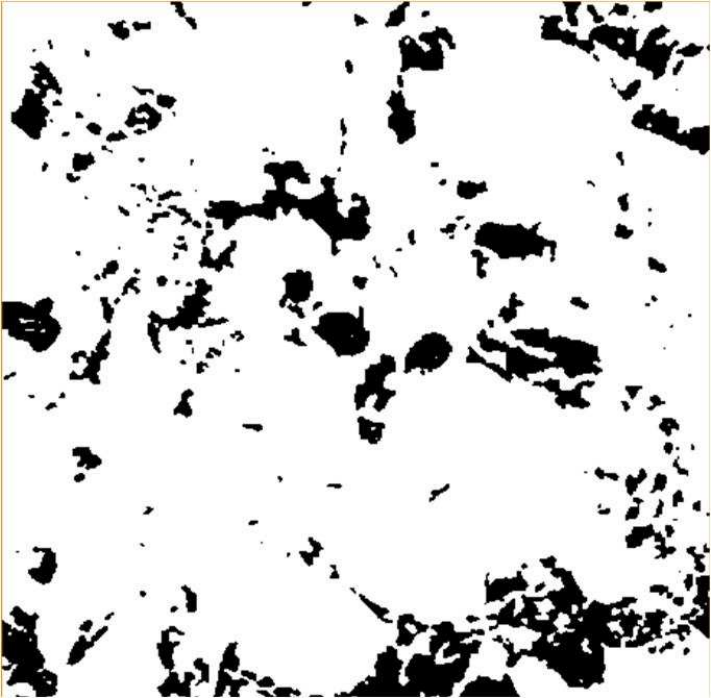}
        \caption{}
    \end{subfigure}
    \hspace{0.2em}
    \begin{subfigure}{.20\textwidth}
        \centering
        \includegraphics[width=\textwidth]{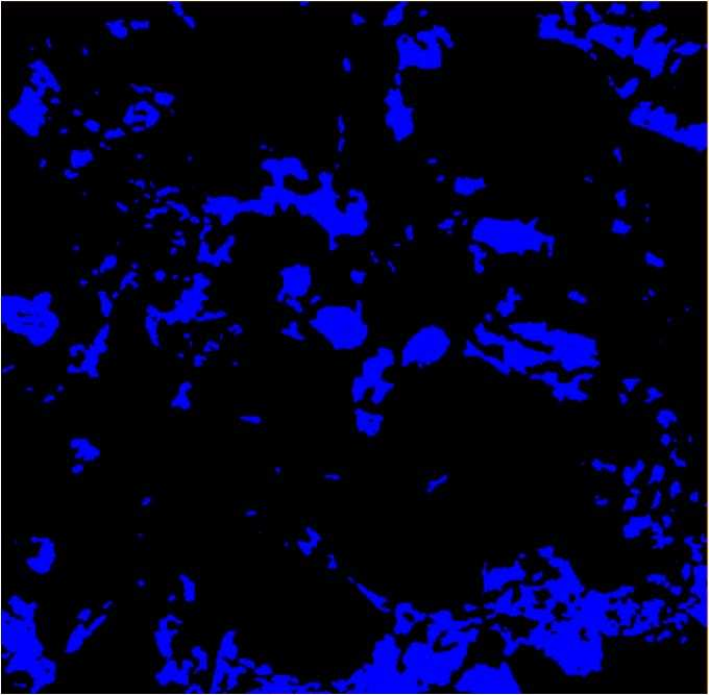}
        \caption{}
    \end{subfigure}
    
    \begin{subfigure}{.20\textwidth}
        \centering
        \includegraphics[width=\textwidth]{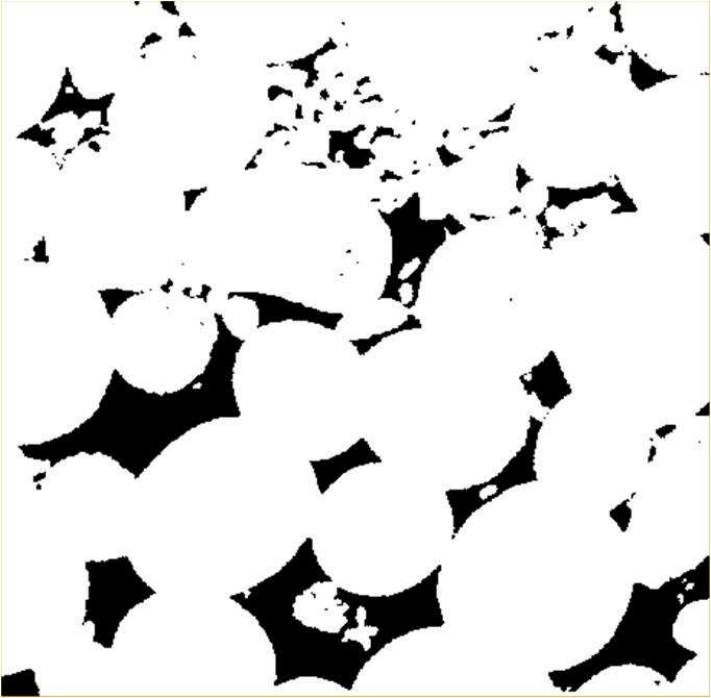}
        \caption{}
    \end{subfigure}
    \hspace{0.2em}
    \begin{subfigure}{.20\textwidth}
        \centering
        \includegraphics[width=\textwidth]{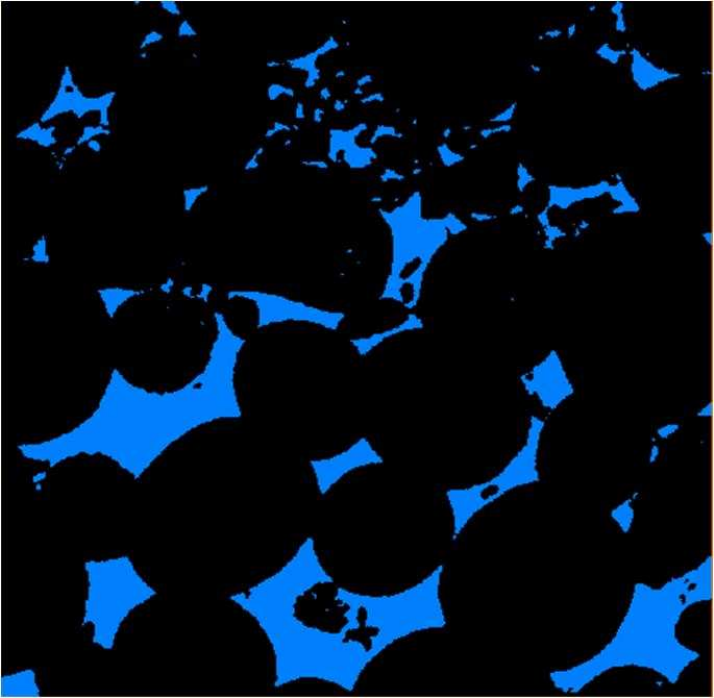}
        \caption{}
    \end{subfigure}

    \caption{Binary CT images and segmented images, where blue areas are pore spaces and black areas are grain spaces. (a) and (b) Bentheimer sandstone; (c) and (d) Doddington sandstone; (e) and (f) Estaillades carbonate; (g) and (h) Ketton carbonate.}
    \label{fig:Binary_Images}
\end{figure}

\begin{figure}[!ht]
  \centering
  \begin{subfigure}{.44\textwidth}  
    \centering
    \includegraphics[width=0.8\textwidth,trim={6cm 3.5cm 6cm 4cm},clip]{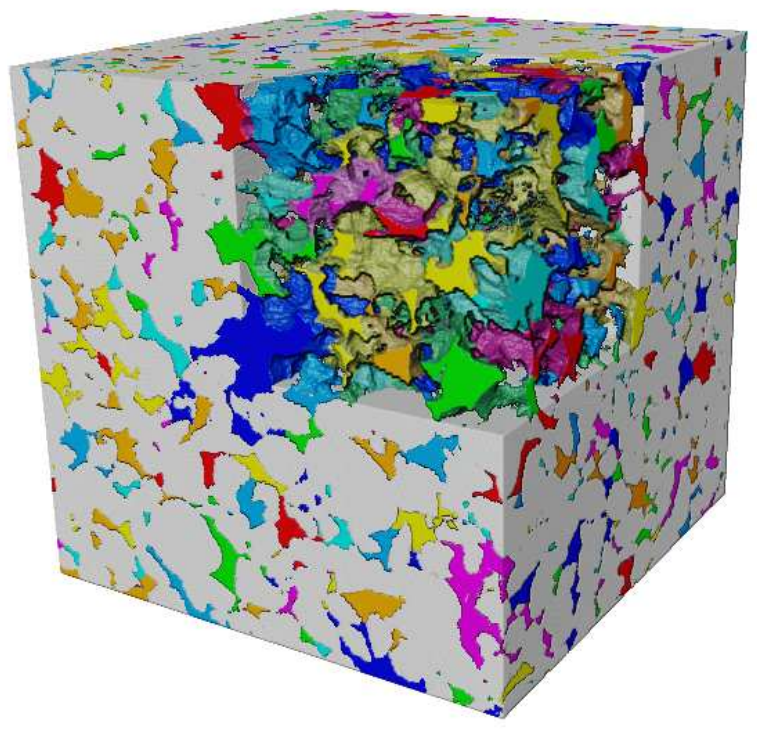}
    \caption{}
  \end{subfigure}
  \hspace{-0.5em}
  \begin{subfigure}{.48\textwidth}
    \centering
    \includegraphics[width=0.8\textwidth,trim={6cm 4cm 6cm 4cm},clip]{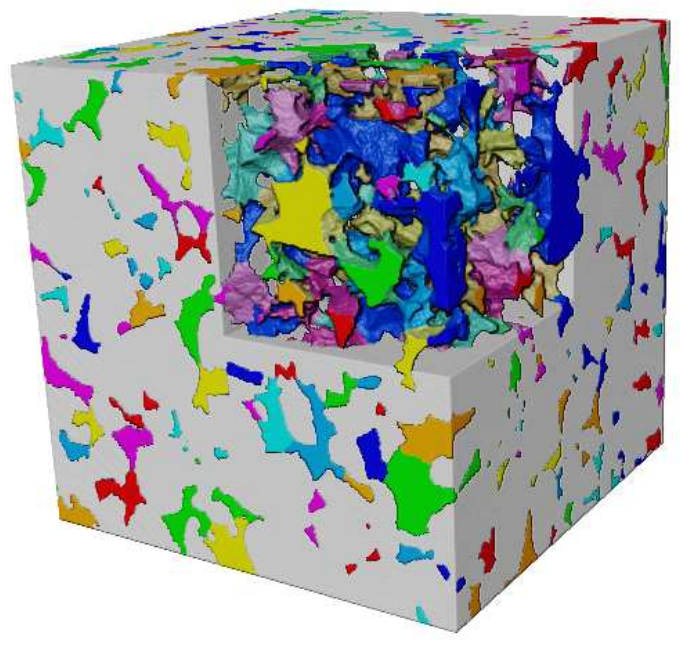}
    \caption{}
  \end{subfigure}
  
  \vspace{1em} 
  \begin{subfigure}{.49\textwidth}
    \centering
    \includegraphics[width=0.8\textwidth,trim={6cm 4cm 6cm 4cm},clip]{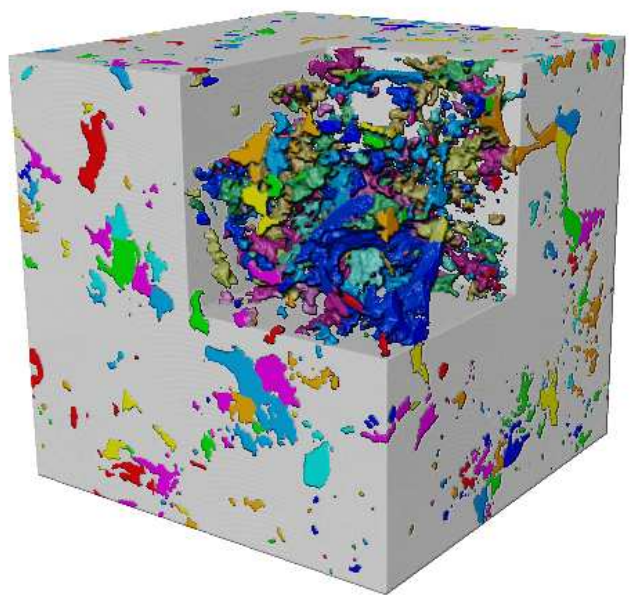}
    \caption{}
  \end{subfigure}
  \hspace{-0.5em}
  \begin{subfigure}{.49\textwidth}
    \centering
    \includegraphics[width=0.8\textwidth,trim={6cm 4cm 6cm 4cm},clip]{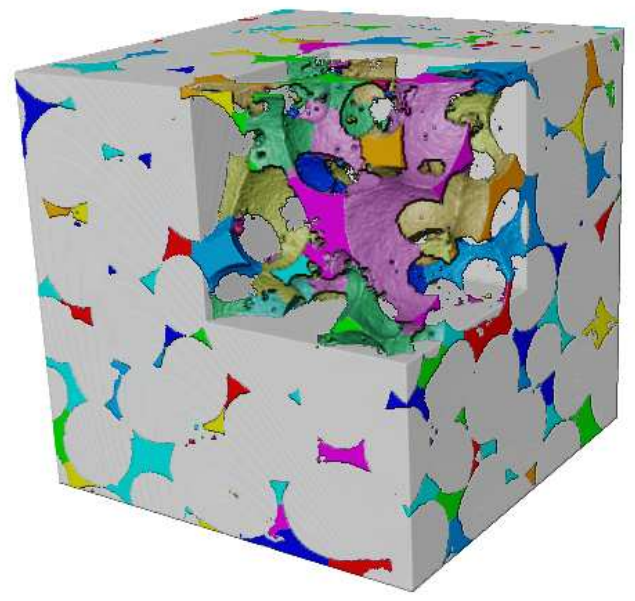}
    \caption{}
  \end{subfigure}
  
  \caption{Reconstructed 3D material domains and pore systems based on CT images: (a) Bentheimer sandstone; (b) Doddington sandstone; (c) Estaillades carbonate; (d) Ketton carbonate.}
  \label{fig:reconstructed images}
\end{figure}

\noindent We use the microCT images of DS, EC and KC (provided by the Imperial College London). BS image was provided by \href{http://www.irocktech.com.cn/en/}{iRock Technologies}. It is noted that the imaged samples are different from the samples used for measuring the bulk properties. 
The raw data provided for each rock is an 8-bit unsigned image, where each voxel is represented by an 8-bit unsigned integer between 0 to 255. The sizes of all images are 1000 voxels in each spatial direction. We used Avizo 3D software (Thermo Fisher Scientific, USA) to perform image segmentation and microstructure characterisation. To avoid boundary effects, the images were cropped to 800 voxels in each direction. The segmentation of voids/pores and grains/particles was performed by the Avizo watershed segmentation algorithm with Gaussian filter and thresholding function. The segmented images are shown in \Cref{fig:Binary_Images}, where the coloured parts are voids. The voxel size and the void volume fraction calculated after segmentation are given in \Cref{tab:rocks_imaging}. A comparison with the porosities in \Cref{tab:rocks_experimental} shows that a significant fraction of the helium-measured porosity in all rocks except the Bentheimer sandstone is not detectable with these voxel resolutions. This suggests that the permeability predictions of models constructed by data from the available images should not be compared with the experimentally measured bulk permeabilities. For this reason, comparisons will be made with fluid dynamics simulations in the imaged void spaces. The reconstructed material domains with coloured voids for the four rocks are shown in \Cref{fig:reconstructed images}. Different colours represent different ordinal numbers of voids. It can be observed in \Cref{tab:rocks_experimental} and \Cref{fig:reconstructed images} that all rocks contain voluminous grains, and large voluminous as well as narrow expansive voids. The distribution of volumes and shapes requires a quantitative analysis.

\setlength{\abovecaptionskip}{0pt}  
\setlength{\belowcaptionskip}{0pt}  

\begin{table}[!ht]
  \caption{Segmentation data}
  \label{tab:rocks_imaging}
  \centering
  \begin{tabular}{ccc}
    \hline
    \noalign{\vskip .25em}
    Material & Voxel size ($\mu m$) & Resolved porosity (\%) \\
    \hline
    \noalign{\vskip .25em}
    Bentheimer sandstone & 3.0 & 21.4 \\[.25em]
    Doddington sandstone & 2.7 & 19.4 \\[.25em]
    Estaillades carbonate & 3.3 & 12.4 \\[.25em]
    Ketton carbonate & 3.0 & 12.6 \\[.25em]
    \hline
  \end{tabular}
\end{table}

\begin{figure}[!ht]
  \begin{subfigure}{.49\textwidth}
    \centering
    \hspace{-0.1\textwidth}
    \includegraphics[scale=.7]{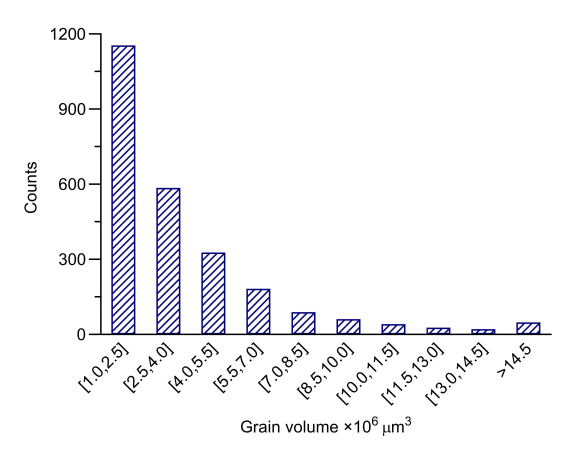}
    \caption{}
  \end{subfigure} 
  \begin{subfigure}{.49\textwidth}
    \centering
    \includegraphics[scale=.7]{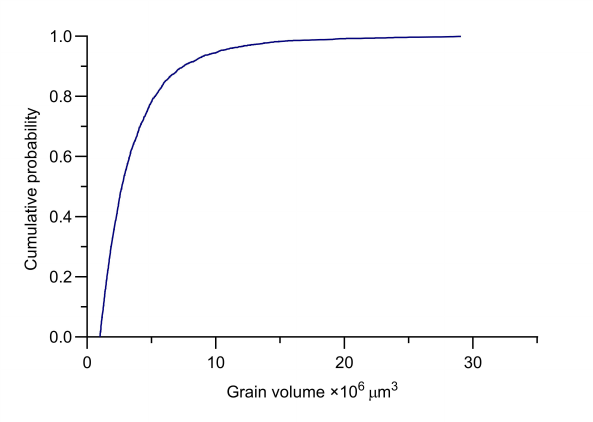}
    \caption{}
  \end{subfigure}
  
  \begin{subfigure}{.49\textwidth}
    \centering
     \hspace{-0.1\textwidth}
    \includegraphics[scale=.7]{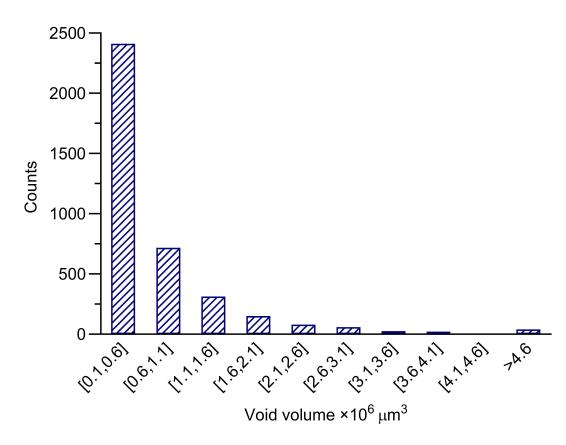}
    \caption{}
  \end{subfigure} 
  \begin{subfigure}{.49\textwidth}
    \centering
    \includegraphics[scale=.7]{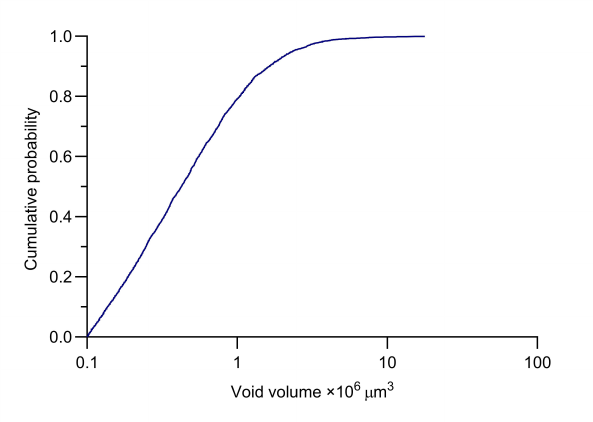}
    \caption{}
  \end{subfigure}

  \caption{Frequency and cumulative probability of resolved grains (a, b) and voids (c, d) in Bentheimer sandstone.}
  \label{fig:BS_statistics}
\end{figure}

The statistical analysis of microstructural features, grains and voids, assumed that a feature had to occupy at least three voxels in each spatial direction for a reliable resolution of its geometry. Therefore features occupying less than 27 voxels were excluded from the statistics; this did not affect grains, but only very small voids in some of the rocks. The resulting statistic, frequency and cumulative probability of grain and void volumes for Bentheimer sandstone is shown in \Cref{fig:BS_statistics}. This rock will be used as an example of the process of model construction. The corresponding statistics for the other three rocks is given in \ref{sec:appendixA}. 

\begin{figure}[!ht]
  \begin{subfigure}{.49\textwidth}
    \centering
    \hspace{-0.1\textwidth}
    \includegraphics[scale=.7]{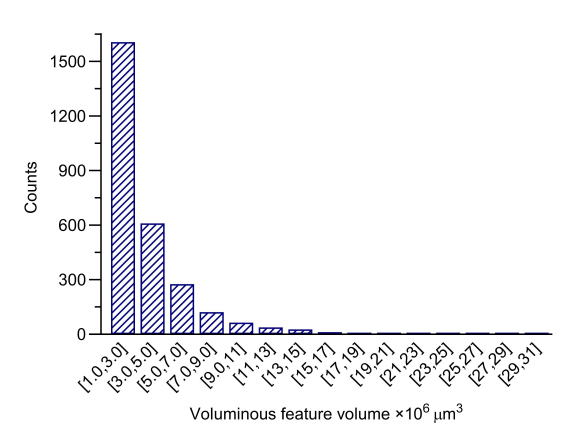}
    \caption{}
    \label{fig:BS_features(a)}
  \end{subfigure} 
  \begin{subfigure}{.49\textwidth}
    \centering
    \includegraphics[scale=.7]{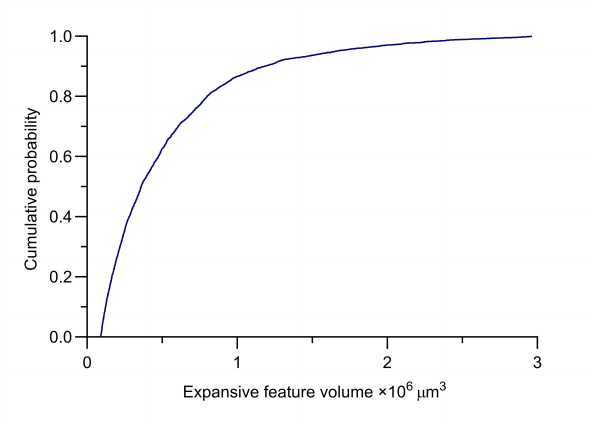}
    \caption{}
    \label{fig:BS_features(b)}
  \end{subfigure}
   \caption{(a) Frequency of voluminous microstructural features (grains and large voids) and (b) cumulative probability of expansive microstructural features (small and large voids) in Bentheimer sandstone.}
  \label{fig:BS_features}
\end{figure}

The microstructural features, grains and voids, were classified using ratios between their spatial extensions determined by Avizo - largest ($L$), intermediate ($I$), and smallest ($S$). Thin long features, characterised by $I/L<0.35$, were not found in the images of the four considered rocks. Voluminous features were characterised by $S/L>0.64$. This criterion classified all grains included in the statistics in \Cref{fig:BS_statistics}, as well as in the other three rocks, as voluminous. In order to keep the range of volumes of voluminous features within reasonable limits for model construction, the voids were first separated into two classes, smaller and larger, than the smallest grain. The voids in the first class were assumed to be expansive, while the voids in the second class were classified into voluminous or expansive based on the same criterion used for grains. The voluminous voids were added to the grain volume frequency to obtain the frequency of voluminous features for model construction, shown in \Cref{fig:BS_features(a)}. The cumulative probability of the remaining voids, considered expansive, is shown in \Cref{fig:BS_features(b)}.

\subsection{Models construction}
\begin{figure}[!ht]
\centering
  \includegraphics[width=.9\linewidth,trim={0cm 1cm 1cm 1cm},clip]{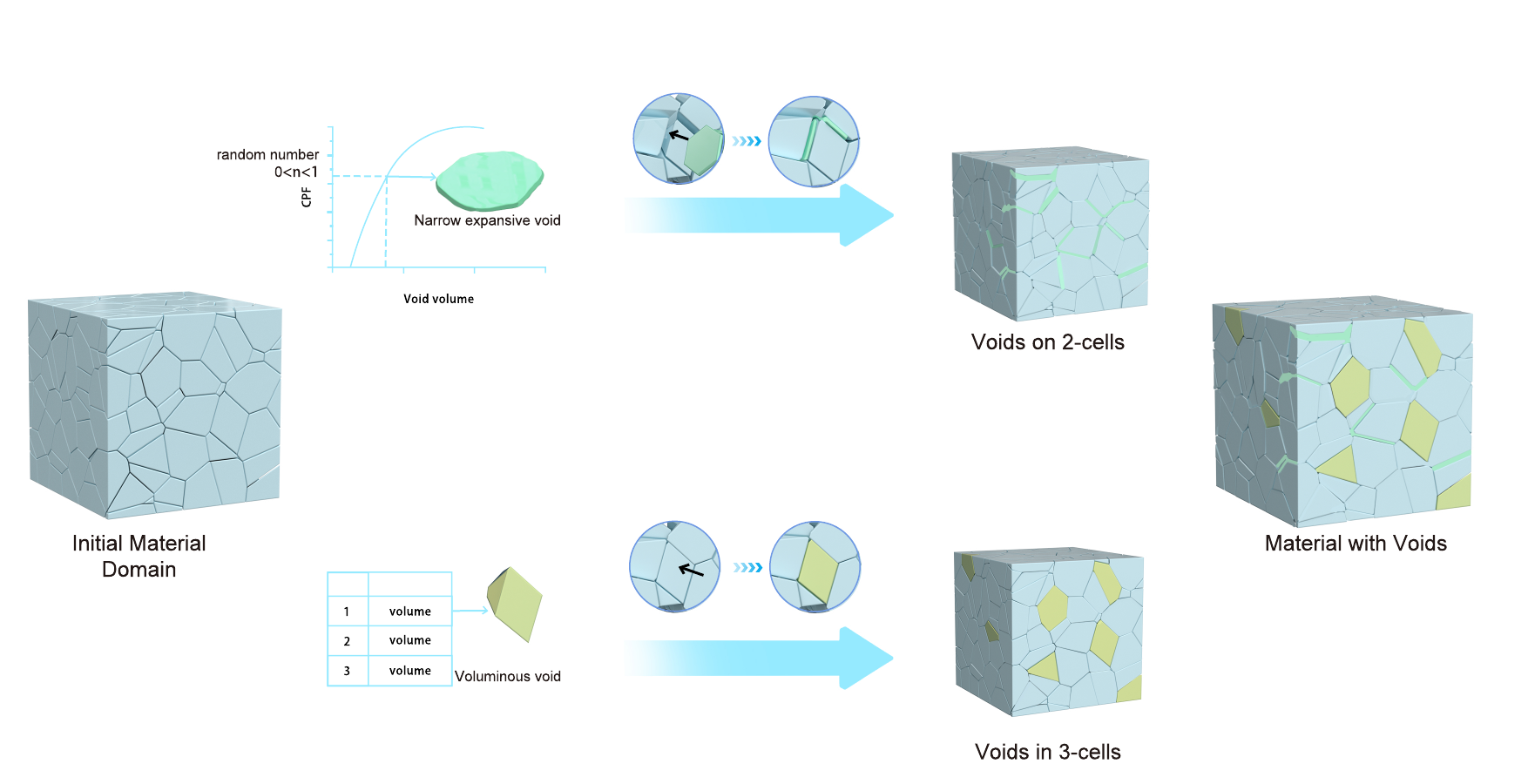}
  \caption{Illustration of pore allocating process.}
  \label{fig:Pore_Allocation}
\end{figure}

\noindent The microstructures of a rock was represented by a cell complex, $\mathcal{M}$, constructed by the open-source software Neper (\url{https://neper.info}), which was developed for modelling polycrystalline assemblies \cite{quey2011large}. In our case, the statistical distribution of volumes of voluminous features, grains and large voluminous voids, shown in \Cref{fig:BS_features(a)}, was given to Neper as input to generate a tessellation of a cubical domain into polyhedral cells with the same volume distribution. The output from Neper was the full geometrical and topological information about the cell complex $\mathcal{M}$. The microstructural features were mapped to the corresponding cells of $\mathcal{M}$ - grains and voluminous voids to $3$-cells, and narrow expansive voids to $2$-cells. A schematic of the process is shown in \Cref{fig:Pore_Allocation}. The mapping of voluminous voids to $3$-cells was performed by selecting $3$-cells with the closest possible volumes, so that the total volume of voids in this class equaled the volume of voids included in \Cref{fig:BS_features(b)}. We note that the $2$-cells on the boundaries of $3$-cells associated with voluminous voids were excluded from subsequent distribution of expansive voids. The mapping of expansive voids to $2$-cells was performed by two random processes. First, a uniform random number $0<p<1$ was generated and the corresponding void volume was determined from \Cref{fig:BS_features(b)}. Second, a $2$-cell was selected at random from the set of $2$-cells with no prior association with an expansive void or a boundary of an voluminous void. The mapping of expansive voids to $2$-cells was terminated when the total volume fraction of all assigned voids equaled the rock's void volume fraction. 

The output was used by our in-house software to generate the computational complex $\mathcal{K}$, whose material properties were assigned as follows. A $3$-cell in $\mathcal{M}$ with mapped pore was declared conductive through the interior, along its $2$-cell boundaries and its $1$-cell boundaries, while a $2$-cell in $\mathcal{M}$ was declared conductive along its surface and along its $1$-cell boundaries. Thus the $1$-cells of $\mathcal{K}$ corresponding to the pairs $(c_0 \prec c_1)$, $(c_1 \prec c_2)$ and $(c_2 \prec c_3)$ of a conductive $3$-cell $c_3$, as well as the $1$-cells of $\mathcal{K}$ corresponding to the pairs $(b_0 \prec b_1)$ and $(b_1 \prec b_2)$ of a conductive $2$-cell $b_2$, were assigned conductivity coefficients for which we used the textbook exact solutions for Poiseuille (pressure-driven) flows \cite{fang2019fluids}. The solution for the velocity $\mathbf{v}$ of a fluid with dynamic viscosity $\mu$ in a cylinder of radius $r$ is 
\begin{equation}
  \mathbf{v} = \frac{r^2}{8\,\mu} \nabla p.
\end{equation}
This was used for calculating the local conductivities of $1$-cells in $\mathcal{K}$ corresponding to the pairs $(b_0 \prec b_1)$, $(c_0 \prec c_1)$, $(c_2 \prec c_3)$ as $\pi= r^2 / (8 \, \mu)$ with $r$ being the half of $h$ in \cref{eq:Conductivity_2cell}, or the half length from midpoint of  $c_1$ to corresponding $c_3$, or the radius of a disk with area equal to the area of $c_2$ respectively. The solution for the velocity $\mathbf{v}$ of a fluid with dynamic viscosity $\mu$ between two plates at distance $h$ is 
\begin{equation}
  \mathbf{v} = \frac{h^2}{12\,\mu} \nabla p.
  \label{eq:Conductivity_2cell}
\end{equation}
This was used for calculating the local conductivities of $1$-cells in $\mathcal{K}$ corresponding to the pairs $(b_1 \prec b_2)$ and $(c_1 \prec c_2)$ as $\pi= h^2 / (12 \, \mu)$. If $b_2$ were a $2$-cell with mapped expansive void, then $h$ was determined by dividing the volume of the mapped void by the area of $b_2$. If $b_2$ were a $2$-cell on the boundary of a $3$-cell $c_3$ with mapped voluminous void, then $h$ was set to the length of the $1$-cell in $\mathcal{K}$ corresponding to the pair $(c_2 \prec c_3)$. All remaining $1$-cells in $\mathcal{K}$ were assigned very small conductivities, $10^{-12} \, \mathrm{m}^2 / (\mathrm{Pa} \cdot \mathrm{s})$, to avoid singularity problems. The dynamic viscosity in all cases was taken to be $1.0 \times 10^{-3}\, {\rm kg} \cdot (m \cdot s)^{-1}$.

\section{Results and discussion}
\label{sec:results_and_discussion}

\begin{figure}[!ht]
  \begin{subfigure}{.49\textwidth}
    \centering
    \hspace{-0.1\textwidth}
    \includegraphics[scale=0.3, trim={2cm 2cm 2cm 2cm}, clip]{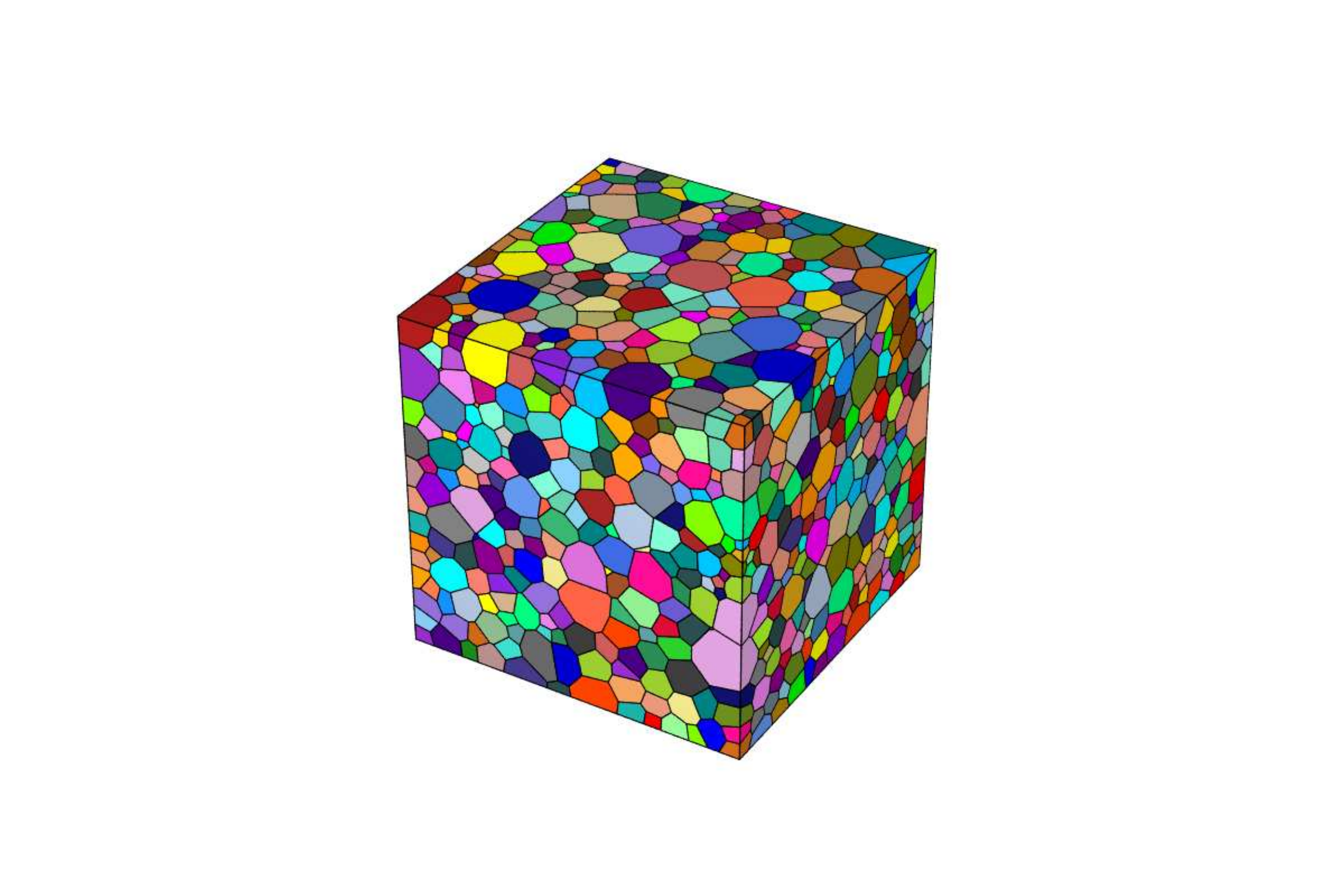}
    \caption{}
  \end{subfigure} 
  \begin{subfigure}{.49\textwidth}
    \centering
    \hspace{-0.1\textwidth}
    \includegraphics[scale=0.3, trim={2cm 2cm 2cm 2cm}, clip]{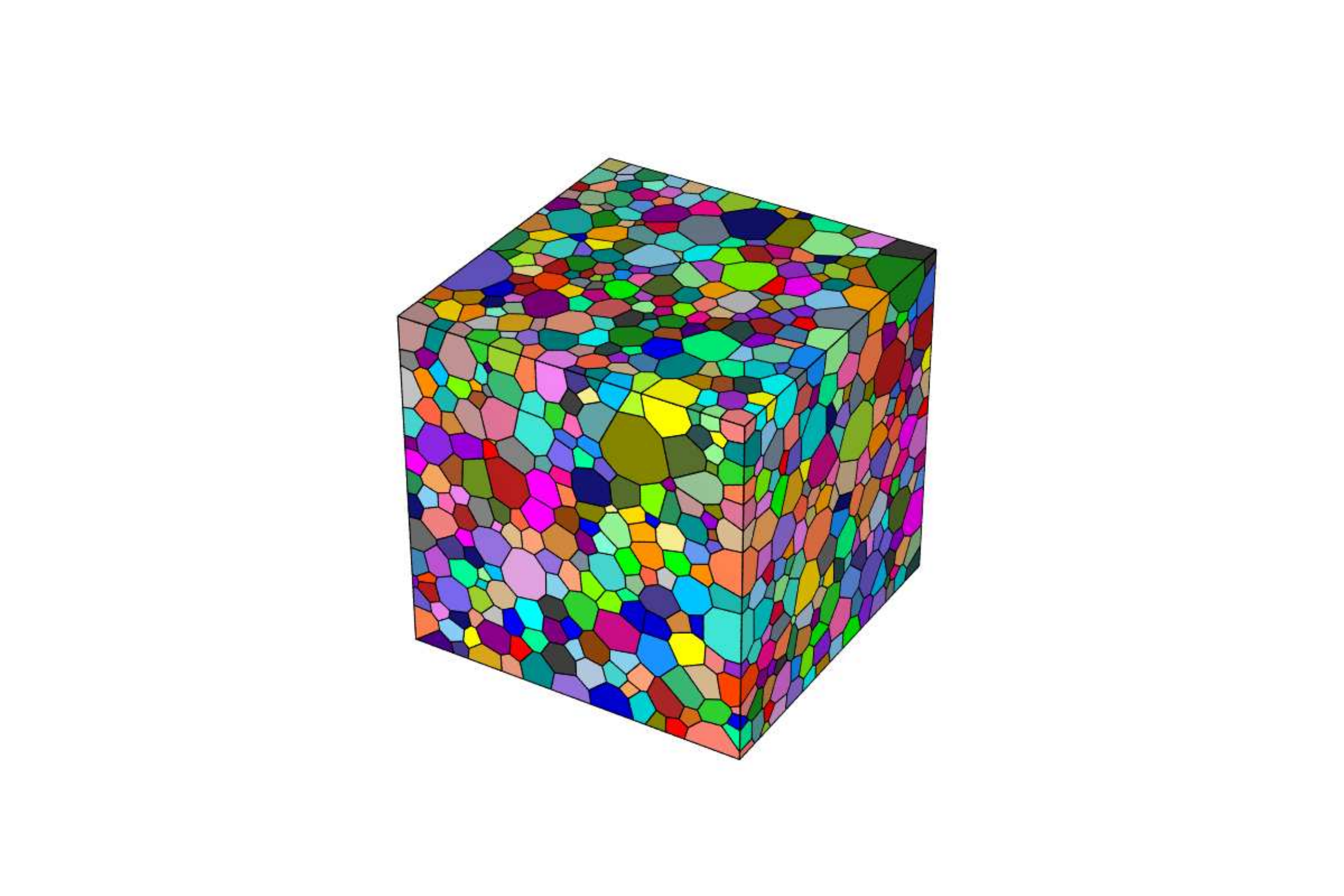}
    \caption{}
  \end{subfigure}
   \caption{Models of Bentheimer sandstone with the same statistical distributions of features' volumes and different features' spatial arrangements.}
  \label{fig:BS_model}
\end{figure}

\begin{figure}[!ht]
  \begin{subfigure}{.49\textwidth}
    \centering
    \hspace{0\textwidth}
    \includegraphics[scale=0.5, trim={2cm 9cm 2cm 9cm}, clip]{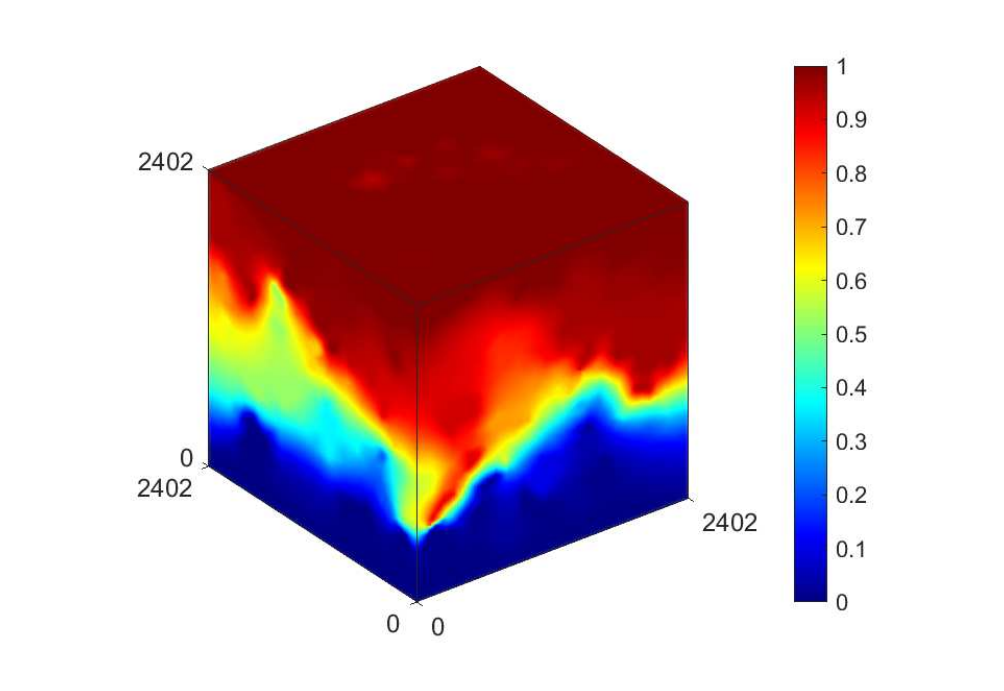}
    \caption{}
  \end{subfigure} 
  \begin{subfigure}{.49\textwidth}
    \centering
    \hspace{0\textwidth}
    \includegraphics[scale=0.5, trim={2cm 9cm 2cm 9cm}, clip]{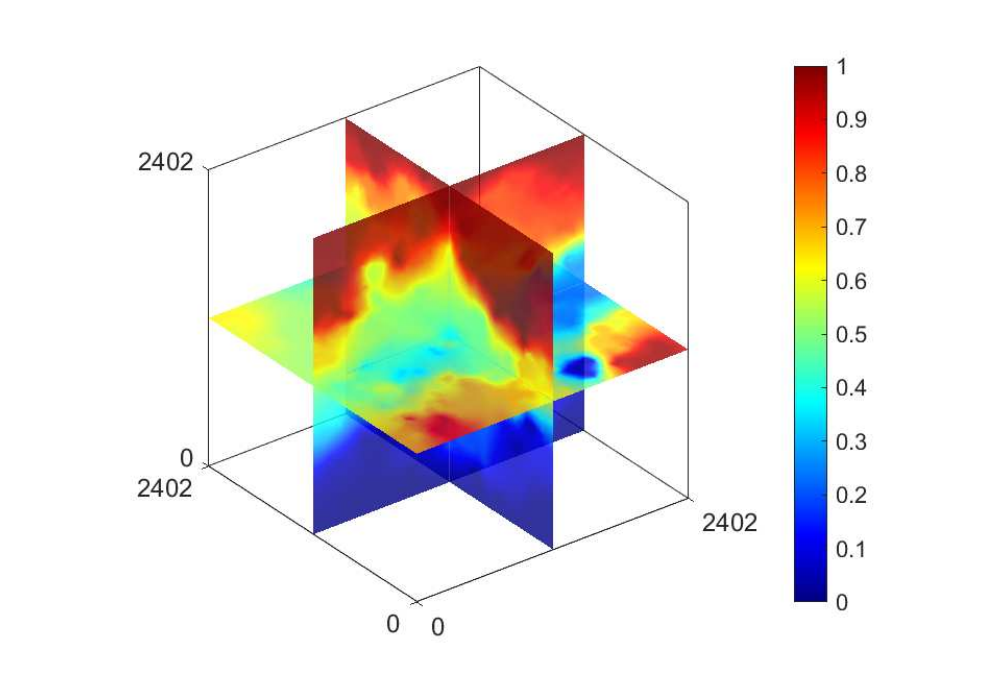}
    \caption{}
  \end{subfigure}
   \caption{Pressure ($\mathrm{Pa}$) distribution in Bentheimer sandstone ($\mu m$)}
  \label{fig:BS_pressure}
\end{figure}

\noindent The statistics of voluminous and expansive features was used to create 30 model realisations for each of the four rocks. An example of two different model realisations with the Bentheimer sandstone's statistics is shown in \Cref{fig:BS_model}. The models were subjected to pressure difference of 1 ($\mathrm{Pa}$) at two opposite faces with the remaining four faces isolated. The pressure distribution calculated with the first realisation at steady-state is shown in \Cref{fig:BS_pressure} to illustrate the heterogeneity of the pressure field dictated by the local conductivities of microstructural features. The plots were created by a natural neighbour interpolation, a three-dimensional algorithm that incorporates neighbouring out-of-plane values into the calculation. As a result, the three planes in the figure provide an in-depth picture of the pressure variation at their intersection and the surroundings. It is worth recalling that pressure values were obtained for all $0$-cells in $\mathcal{K}$. 
The emergent/macroscopic conductivity and hence permeability values in each of the three directions of the cubical models were calculated for each model realisation, thus providing 90 permeability values for each rock. The average permeability values calculated by the proposed modelling and simulation method are given in the second column of \Cref{tab:results}.

In addition, the permeability of the images of the four rocks were calculated by the Absolute Permeability Experiment Simulation, which is a function in XLab Hydro module of Avizo. The calculations in this simulator are realized by solving Navier-Stokes equations in the void space and using Darcy’s law to obtain the permeability. The flow equations are solved with the assumptions of incompressible fluid with constant dynamic viscosity (the same as the value used in our models), and laminar flow (the same as the assumption for the local conductivities in our models). The boundary conditions were the same as for the new method with a no-slip condition at fluid-solid interfaces. The boundary pressure difference was set at 1 ($\mathrm{Pa}$) as for the calculations with the new model. The permeability values calculated by Avizo are given in the third column of \Cref{tab:results}.

For (partial) validation we include in the fourth column of \Cref{tab:results} the experimentally measured permeability values \cite{raeini2017generalized,andrew2014reservoir}, noting again that these were obtained with samples of the four rocks larger than the imaged samples. The measurements of Estaillades carbonate, Doddington sandstone and Ketton carbonate were conducted at Weatherford Laboratories (East Grinstead, UK) and the measurement of Bentheimer sandstone were conducted at Imperial College (London, UK). In all cases the samples were cylinders with 6.5 ${\rm mm}$ diameter and 22 ${\rm mm}$ height. 

\begin{table}[!ht]
  \caption{A summary of results for all rock samples}
  \label{tab:results}
  \centering
  \begin{tabular}{cccc}
    \hline
    \noalign{\vskip .25em}
    Material & \parbox{5cm}{\centering Average permeability \\ from 30 model realisations \\ ($10^{-12} \, m^2$)} & \parbox{4cm}{\centering Single permeability \\ computed by CFD \\ ($10^{-12} \, m^2$)} & \parbox{4cm}{\centering Single experimental \\ permeability \\ ($10^{-12} \, m^2$)} \\
    \hline
    \noalign{\vskip .25em}
    Bentheimer sandstone & 4.32 & 6.89 &  1.875 \\[.25em]
    Doddington sandstone & 3.84 & 6.08 & 1.038 \\[.25em]
    Estaillades carbonate &  1.59& 2.11 & 0.149 \\[.25em]
    Ketton carbonate & 4.47 & 5.38 & 2.807 \\[.25em]
    \hline
  \end{tabular}
\end{table}

\begin{figure}[!ht]
  \centering
  \includegraphics[scale=1,trim={0cm 0cm 0cm 0cm},clip]{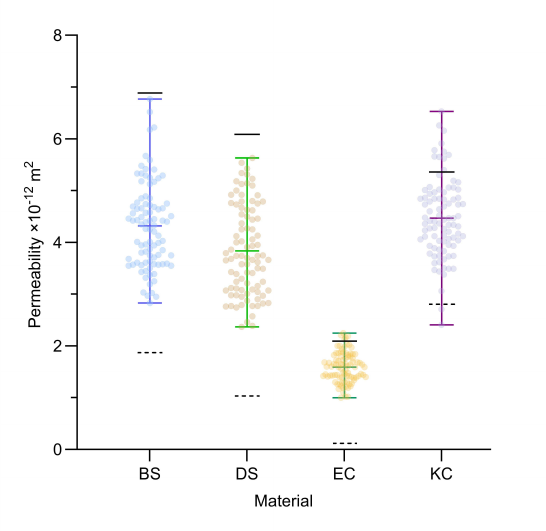}
  \caption{Full set of permeability values of the four rocks}
  \label{fig:result}
\end{figure}

The results in \Cref{tab:results} show that the new modelling and simulation method delivers permeability values in good agreement with the classical CFD and with experiments. \Cref{fig:result} shows the full sets of calculated permeability values, depicted by dots, the CFD computed values with black solid line, and the measured values with black dashed line for all studied rocks. These results demonstrate two things. First, a comparison of the CFD and experimental values show that the fabrics (statistics and spatial arrangement of voids and grains volumes and shapes) of the segmented images were different from the fabrics of the specimens used for experimental measurement of permeability. Secondly, the proposed modelling approach can take the statistics of grains and voids volumes and shapes and by testing different arrangements predict the spread of the macroscopic property values that is expected for given statistics. The observation that in most cases the experimental values are not within the calculated spread suggest that the statistics extracted from the imaged samples are not the same as for the experimental samples. Improving the statistics, i.e., considering a larger set of imaged samples from the rocks is expected to improve the predictive capability. 

It is noted that the local conductivity values assigned to different cells of the cell complex are taken from the solutions of flow in cylindrical and ractangular channels, which is a simplification of the real pore-scale morphologies. One possible approach to account for real morphologies is to consider the void surfaces as having fractal dimensions.

An important advantage of the proposed method is its computational efficiency. We show in \Cref{tab:resources} the average computational resources for a single simulation with the new method and the CFD simulations for the four rocks. It should be noted when making the comparison, that the new method was used on a single PC with a working code written in Matlab, which can be found at \url{https://github.com/changhaoliu-UOM/flow_2024.git}, while the CFD was performed on a GPU array, which provides further support for the efficiency of the new formulation.

\begin{table}[!ht]
  \caption{Computational resources required for a single simulation}
  \label{tab:resources}
  \centering
  \begin{tabular}{ccc}
    \hline
    \noalign{\vskip .25em}
    Material & \parbox{5cm}{\centering Memory and CPU time} & \parbox{5cm}{\centering Memory and GPU time} \\
    \hline
    \noalign{\vskip .25em}
    Bentheimer sandstone & 988 MB, 209 sec & 50.3 GB, 19 min \\[.25em]
    Doddington sandstone & 445 MB, 78 sec & 51.8 GB, 22 min\\[.25em]
    Estaillades carbonate & 580 MB, 122 sec & 50.6 GB, 21 min \\[.25em]
    Ketton carbonate & 590 MB, 54 sec & 53.1 GB, 24 min\\[.25em]
    \hline
  \end{tabular}
\end{table}

\section{Conclusions}
\noindent We have presented a new foundation for modelling the materials systems where they are represented by cell complexes - collections of cells of different topological dimensions. This is markedly different from the two classical approaches, where materials systems are represented as either finite collections of distinct points (discrete topology) leading to various particle methods, such as molecular dynamics, peridynamics, smoothed particle hydrodynamics, etc., or as infinite collections of infinitesimal points (smooth topology) leading to the continuum description of physics solved by spatial discretisation methods, such as finite elements, finite volumes, finite differences, etc. The advantages of the new modelling foundations are as follows: 
\begin{itemize}
  \item 
    It allows for faithful representation of materials with complex internal structures, where elements with different apparent dimensions may have different material properties and may take part in the physical processes in interaction with others. This feature has been demonstrated in the present work by the construction of models for four rocks using experimentally determined grain and void volume and shape distributions.
  \item
    It permits formulations of the conservation laws of physics that respect the underlying structure - the topology and geometry of the cell complex together with the material properties of its components. Importantly, these formulations are directly in matrix form and are exact for a given complex, i.e., the method uses discrete operators in contrast to classical numerical methods which use discretisation of material domains and approximation of continuum operators. This feature has been demonstrated in the present work by the simulation of fluid flow in the models of four rock to calculate their macroscopic property permeability.
  \item 
    It facilitates investigation of structure-properties relations in materials by linking topological characteristics of sub-structures, e.g., the sub-structures of voids and grains, to calculated macroscopic properties. This feature has not been explored in the present work and is part of ongoing research. Specifically for porous media, such investigation is expected to demonstrate whether the permeability is a function only of the three classical bulk parameters - porosity, tortuosity, and grain surface area - or further parameter(s) are required to capture differences in the arrangement of fabric's features.
\end{itemize}

\noindent The models construction and solutions for flow have shown that:
\begin{itemize}
  \item 
    The tessellation of domains into polyhedrons using the statistics of voluminous features by Neper creates suitable 3D canvases for introducing all observed features of the materials' fabrics in a sufficiently realistic manner. The algorithm for mapping observed features to elements of the cell complex produces heterogeneous models with specified void and grain volume distributions that can be used for studying the effect of different arrangements of features on the emergent property.
  \item
    The calculated permeability values are in good agreement with the CFD calculations and the experimentally measured values for all rocks. The calculated spread of values with different fabrics' arrangements shows that the proposed modelling and simulation method can be used to estimate the variability of a macroscopic property for a given statistics and that the estimation can be improved by improving the input statistics.
  \item
    The proposed modelling and simulation method requires an order of magnitude less time and two orders of magnitude less memory that the classical CFD for calculating the macroscopic property permeability.
\end{itemize}

\noindent Importantly, the proposed approach can be used to investigate the evolution of macroscopic properties due to changes in the fabric in a simple and effective way. For example, a non-linear version of the solution method will allow to analyse flow in evolving pore spaces because of different factors, such as dissolution and precipitation, deformation, and fracture propagation. The mathematical formulation is identical for all conservation laws involving scalar potentials (mass density, temperature, electric charge), which allows for seamless coupling of different processes, e.g., fluid flow, mass diffusion, and heat conduction, with different material properties and couplings in different elements of the fabric. The conservation laws for linear and angular momenta, needed for analysis of deformation and fracture, involve (classically) vector potentials, displacement or velocity fields. The discrete versions of these laws require the introduction of discrete bundle-valued differential forms. The development of the corresponding apparatus is analogous to the recently published excellent work on elasticity with smooth exterior calculus \cite{rashad2023exteriorelasticity}, and is a subject of ongoing work. 


\appendix

\section{Statistics of resolved features in rocks}
\label{sec:appendixA}
\begin{figure}[!ht]
  \begin{subfigure}{.49\textwidth}
    \centering
    \hspace{-0.1\textwidth}
    \includegraphics[scale=.7]{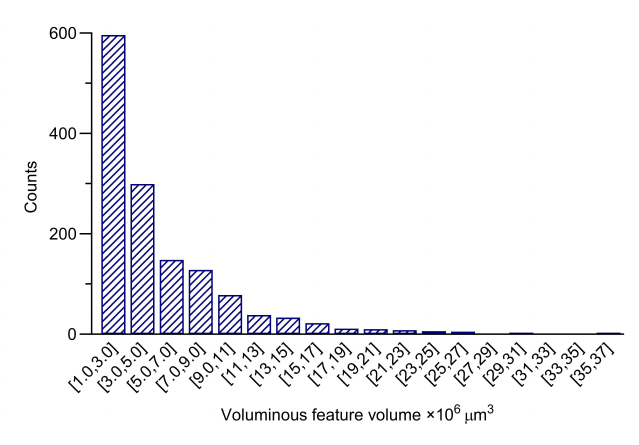}
    \caption{}
  \end{subfigure} 
  \begin{subfigure}{.49\textwidth}
    \centering
    \includegraphics[scale=.7]{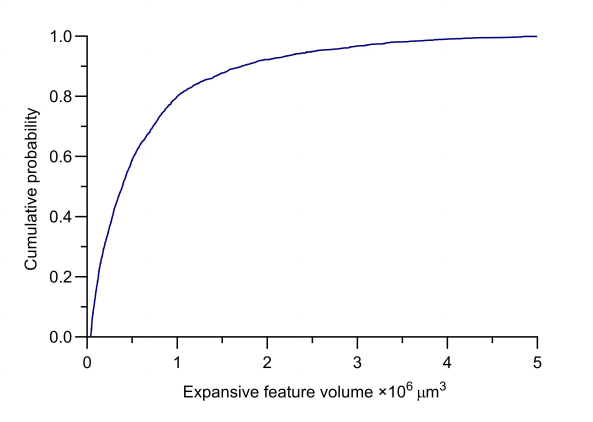}
    \caption{}
  \end{subfigure}
   \caption{(a) Frequency of voluminous microstructural features (grains and large voids) and (b) cumulative probability of expansive microstructural features (small and large voids) in Doddington sandstone}
  \label{fig:DS_features}
\end{figure}

\begin{figure}[!ht]
  \begin{subfigure}{.5\textwidth}
    \centering
     \hspace{-0.1\textwidth}
    \includegraphics[scale=.7]{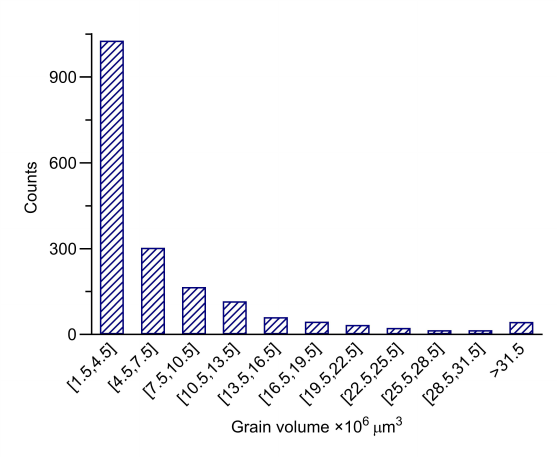}
    \caption{}
  \end{subfigure} 
  \begin{subfigure}{.5\textwidth}
    \centering
    \includegraphics[scale=.7]{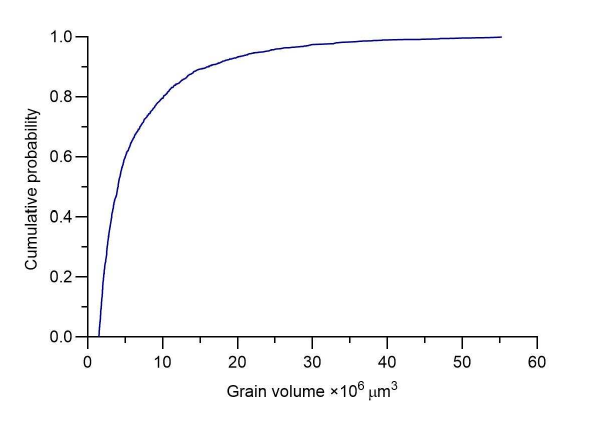}
    \caption{}
  \end{subfigure}
  
  \begin{subfigure}{.5\textwidth}
    \centering
     \hspace{-0.1\textwidth}
    \includegraphics[scale=.7]{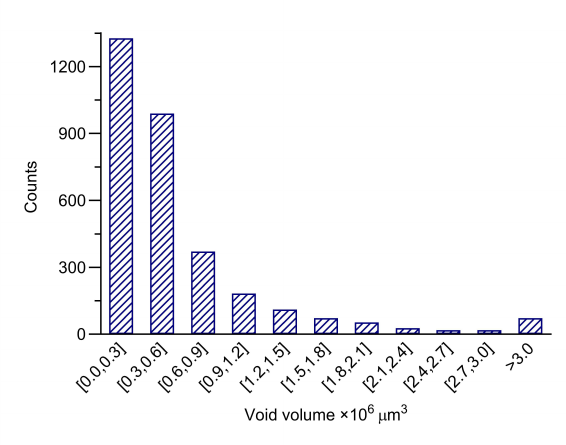}
    \caption{}
  \end{subfigure} 
  \begin{subfigure}{.5\textwidth}
    \centering
    \includegraphics[scale=.7]{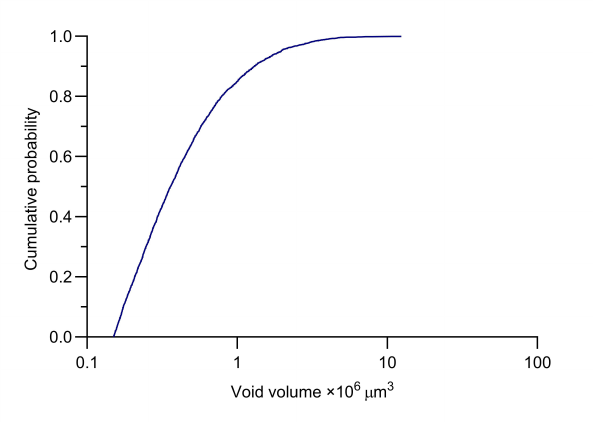}
    \caption{}
  \end{subfigure}
  \caption{Frequency and cumulative probability of resolved grains (a, b) and voids (c, d) in Estaillades carbonate}
  \label{fig:EC_statistics}
\end{figure}

\begin{figure}[!ht]
  \begin{subfigure}{.49\textwidth}
    \centering
    \hspace{-0.1\textwidth}
    \includegraphics[scale=.7]{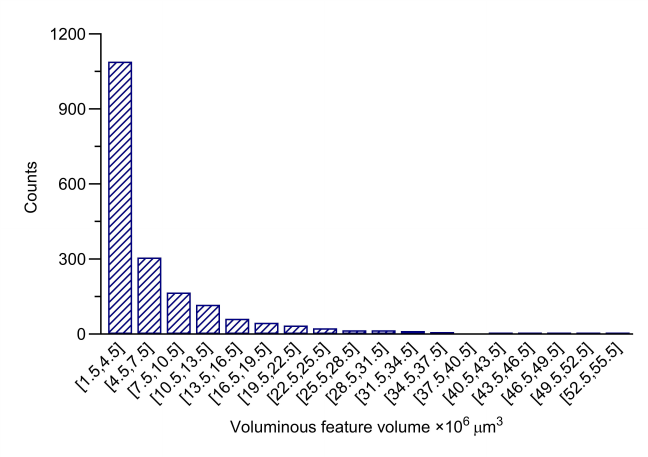}
    \caption{}
  \end{subfigure} 
  \begin{subfigure}{.49\textwidth}
    \centering
    \includegraphics[scale=0.7]{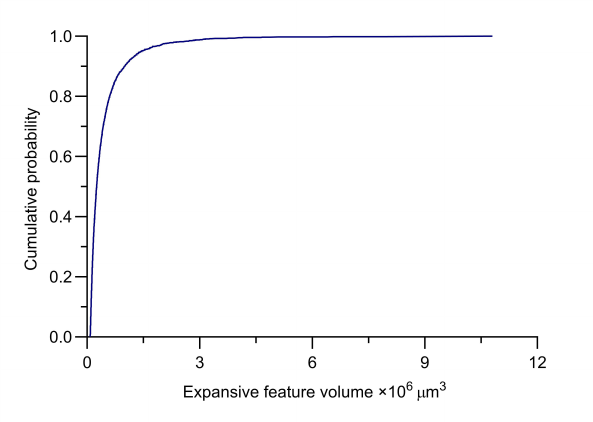}
    \caption{}
  \end{subfigure}
   \caption{(a) Frequency of voluminous microstructural features (grains and large voids) and (b) cumulative probability of expansive microstructural features (small and large voids) in Estaillades carbonate}
  \label{fig:EC_features}
\end{figure}

\begin{figure}[!ht]
  \begin{subfigure}{.5\textwidth}
    \centering
     \hspace{-0.2\textwidth}
    \includegraphics[scale=.7]{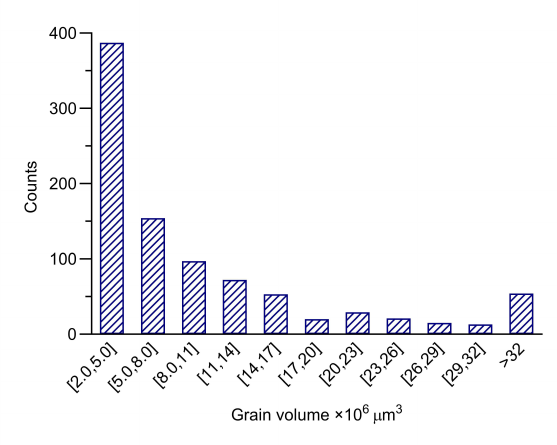}
    \caption{}
  \end{subfigure} 
  \begin{subfigure}{.5\textwidth}
    \centering
    \includegraphics[scale=.7]{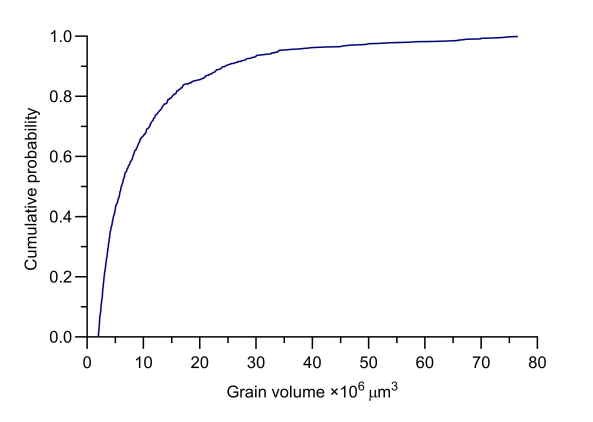}
    \caption{}
  \end{subfigure}
  
  \begin{subfigure}{.5\textwidth}
    \centering
     \hspace{-0.2\textwidth}
    \includegraphics[scale=.7]{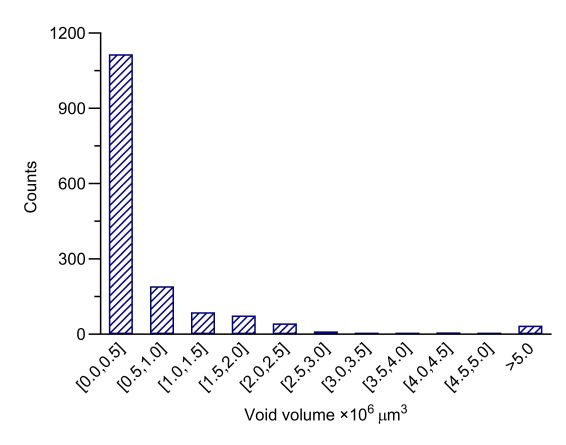}
    \caption{}
  \end{subfigure} 
  \begin{subfigure}{.5\textwidth}
    \centering
    \includegraphics[scale=.7]{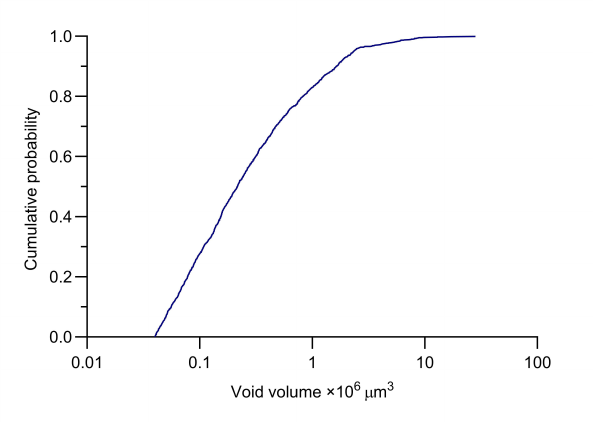}
    \caption{}
  \end{subfigure}
  \caption{Frequency and cumulative probability of resolved grains (a, b) and voids (c, d) in Ketton carbonate}
  \label{fig:KC_statistics}
\end{figure}

\begin{figure}[!ht]
  \begin{subfigure}{.49\textwidth}
    \centering
    \hspace{-0.1\textwidth}
    \includegraphics[scale=0.7]{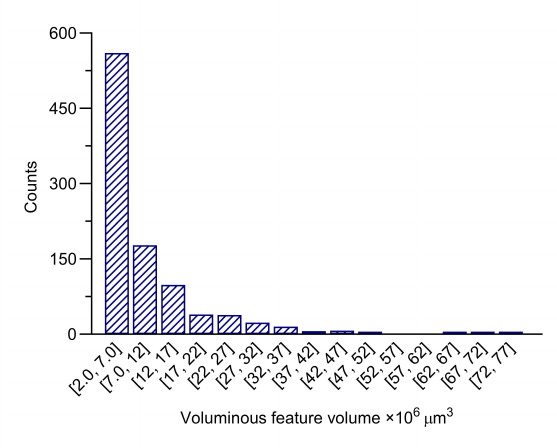}
    \caption{}
  \end{subfigure} 
  \begin{subfigure}{.49\textwidth}
    \centering
    \includegraphics[scale=.7]{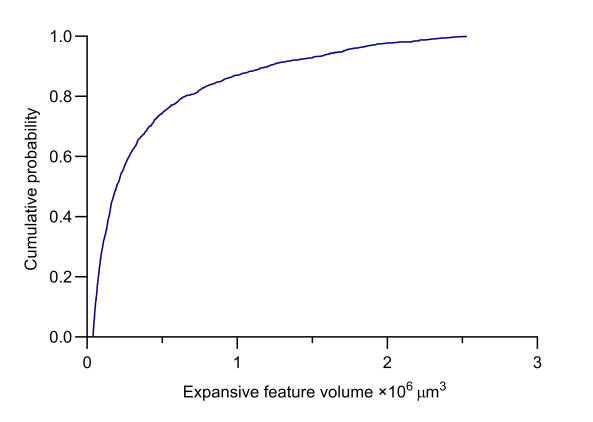}
    \caption{}
  \end{subfigure}
   \caption{(a) Frequency of voluminous microstructural features (grains and large voids) and (b) cumulative probability of expansive microstructural features (small and large voids) in Ketton carbonate}
  \label{fig:KC_features}
\end{figure}

\noindent The statistics of the resolved microstructural features and those used for constructing the models of the remaining three rocks are shown in the following figures: Doddington sandstone (\Cref{fig:DS_statistics}, \Cref{fig:DS_features}), Estaillades carbonate (\Cref{fig:EC_statistics}, \Cref{fig:EC_features}) and Ketton carbonate (\Cref{fig:KC_statistics}, \Cref{fig:KC_features}).
\begin{figure}[!ht]
  \begin{subfigure}{.5\textwidth}
    \centering
     \hspace{-0.1\textwidth}
    \includegraphics[scale=.7]{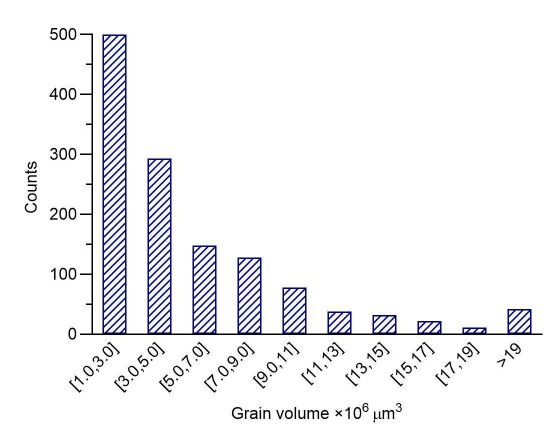}
    \caption{}
  \end{subfigure} 
  \begin{subfigure}{.5\textwidth}
    \centering
    \includegraphics[scale=.7]{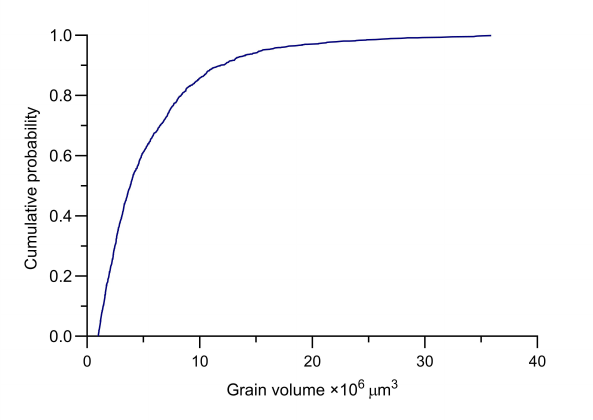}
    \caption{}
  \end{subfigure}
  
  \begin{subfigure}{.5\textwidth}
    \centering
     \hspace{-0.1\textwidth}
    \includegraphics[scale=.7]{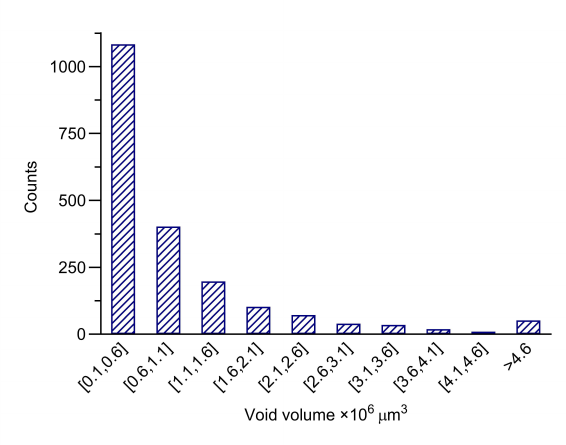}
    \caption{}
  \end{subfigure} 
  \begin{subfigure}{.5\textwidth}
    \centering
    \includegraphics[scale=.7]{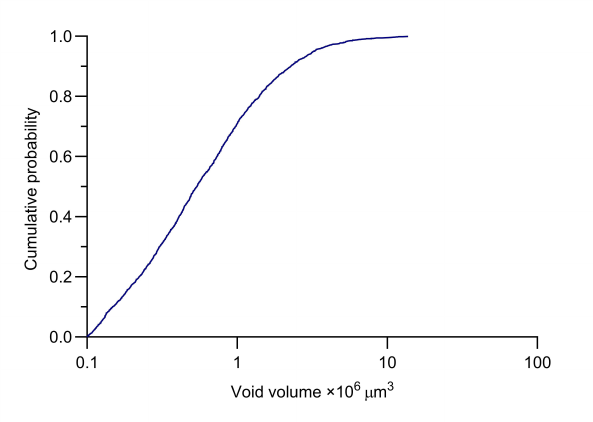}
    \caption{}
  \end{subfigure}
  \caption{Frequency and cumulative probability of resolved grains (a, b) and voids (c, d) in Doddington sandstone}
  \label{fig:DS_statistics}
\end{figure}

\clearpage

\section{Examples of model realisations and pressure solutions for rocks}
\label{appendixB}
\begin{figure}[!ht]
  \begin{subfigure}{.49\textwidth}
    \centering
    \hspace{-0.1\textwidth}
    \includegraphics[scale=0.2, trim={2cm 2cm 2cm 2cm}, clip]{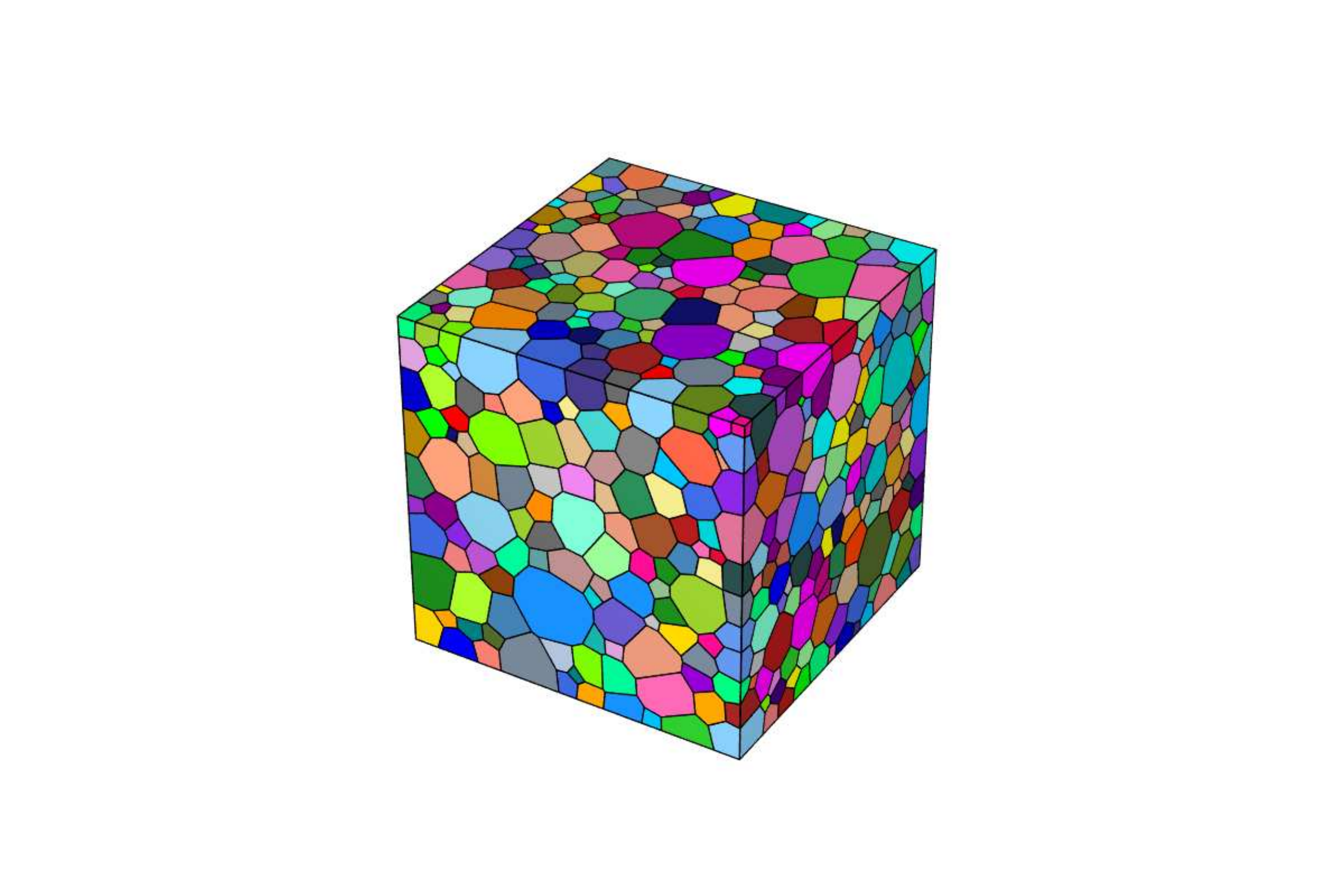}
    \caption{}
  \end{subfigure} 
  \begin{subfigure}{.49\textwidth}
    \centering
    \hspace{-0.1\textwidth}
    \includegraphics[scale=0.2, trim={2cm 2cm 2cm 2cm}, clip]{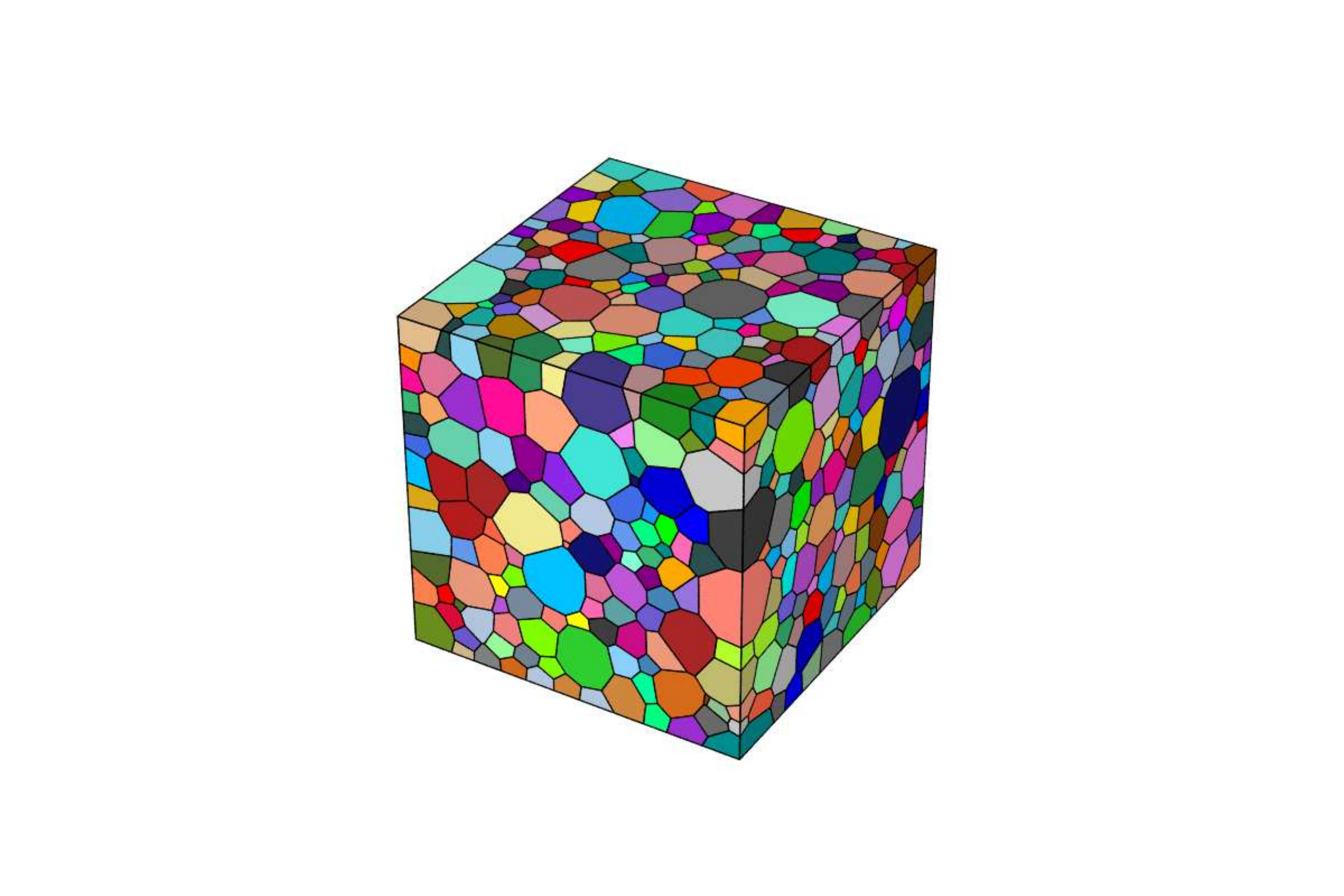}
    \caption{}
  \end{subfigure}
   \caption{Reconstructed cubical meshes of Doddington sandstone}
  \label{fig:DS_model}
\end{figure}

\begin{figure}[!ht]
  \begin{subfigure}{.49\textwidth}
    \centering
    \hspace{0\textwidth}
    \includegraphics[scale=0.4, trim={2cm 9cm 2cm 9cm}, clip]{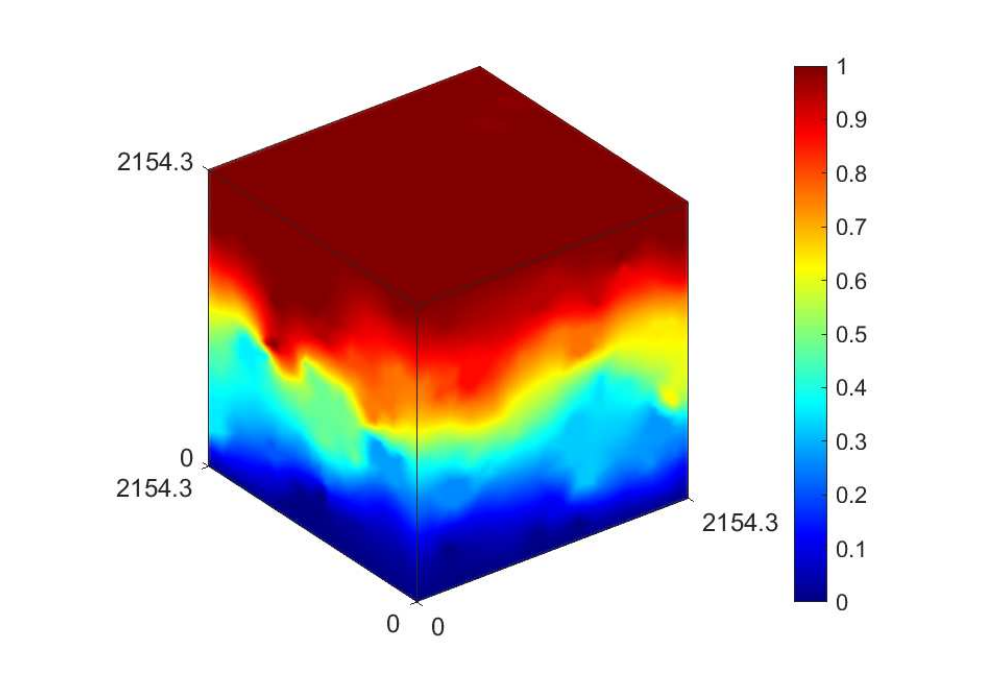}
    \caption{}
  \end{subfigure} 
  \begin{subfigure}{.49\textwidth}
    \centering
    \hspace{0\textwidth}
    \includegraphics[scale=0.4, trim={2cm 9cm 2cm 9cm}, clip]{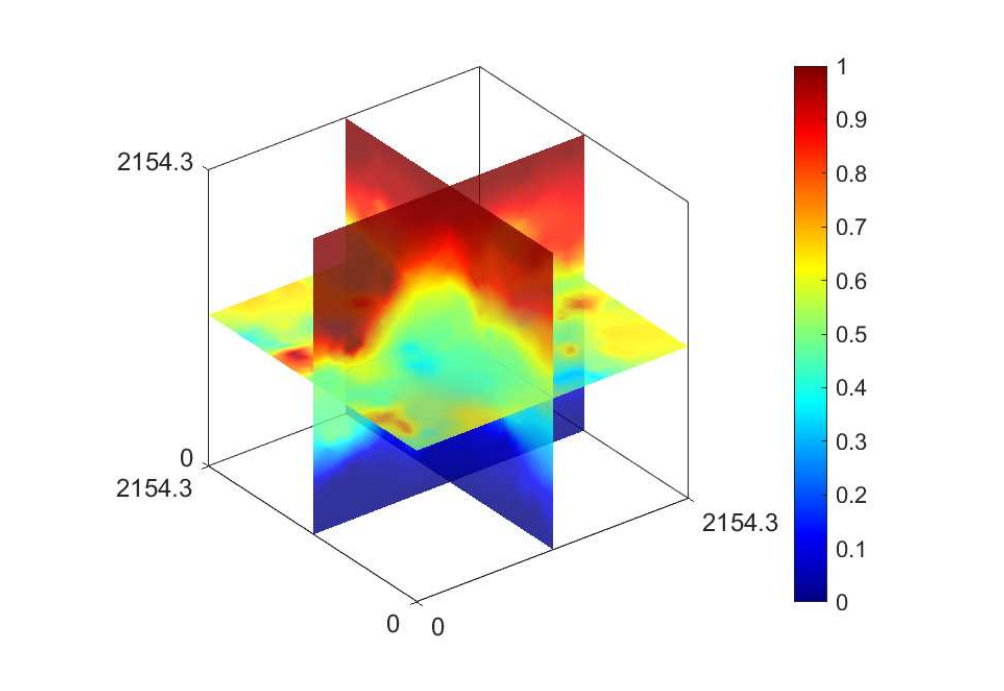}
    \caption{}
  \end{subfigure}
   \caption{Pressure ($\mathrm{Pa}$) distribution in Doddington sandstone ($\mu m$)}
  \label{fig:DS_pressure}
\end{figure}

\begin{figure}[!ht]
  \begin{subfigure}{.49\textwidth}
    \centering
    \hspace{-0.1\textwidth}
    \includegraphics[scale=0.2, trim={2cm 2cm 2cm 2cm}, clip]{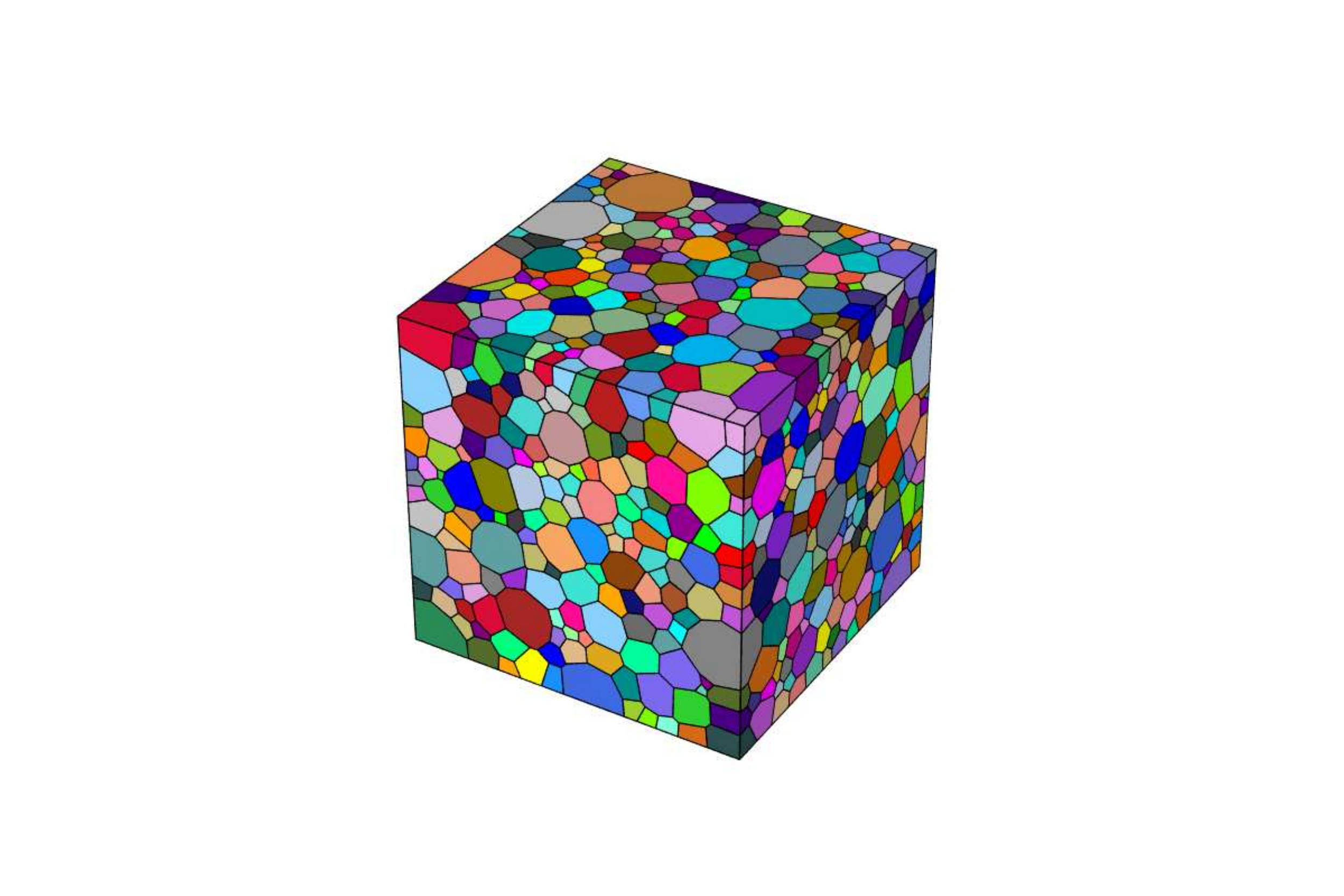}
    \caption{}
  \end{subfigure} 
  \begin{subfigure}{.49\textwidth}
    \centering
    \hspace{-0.1\textwidth}
    \includegraphics[scale=0.2, trim={2cm 2cm 2cm 2cm}, clip]{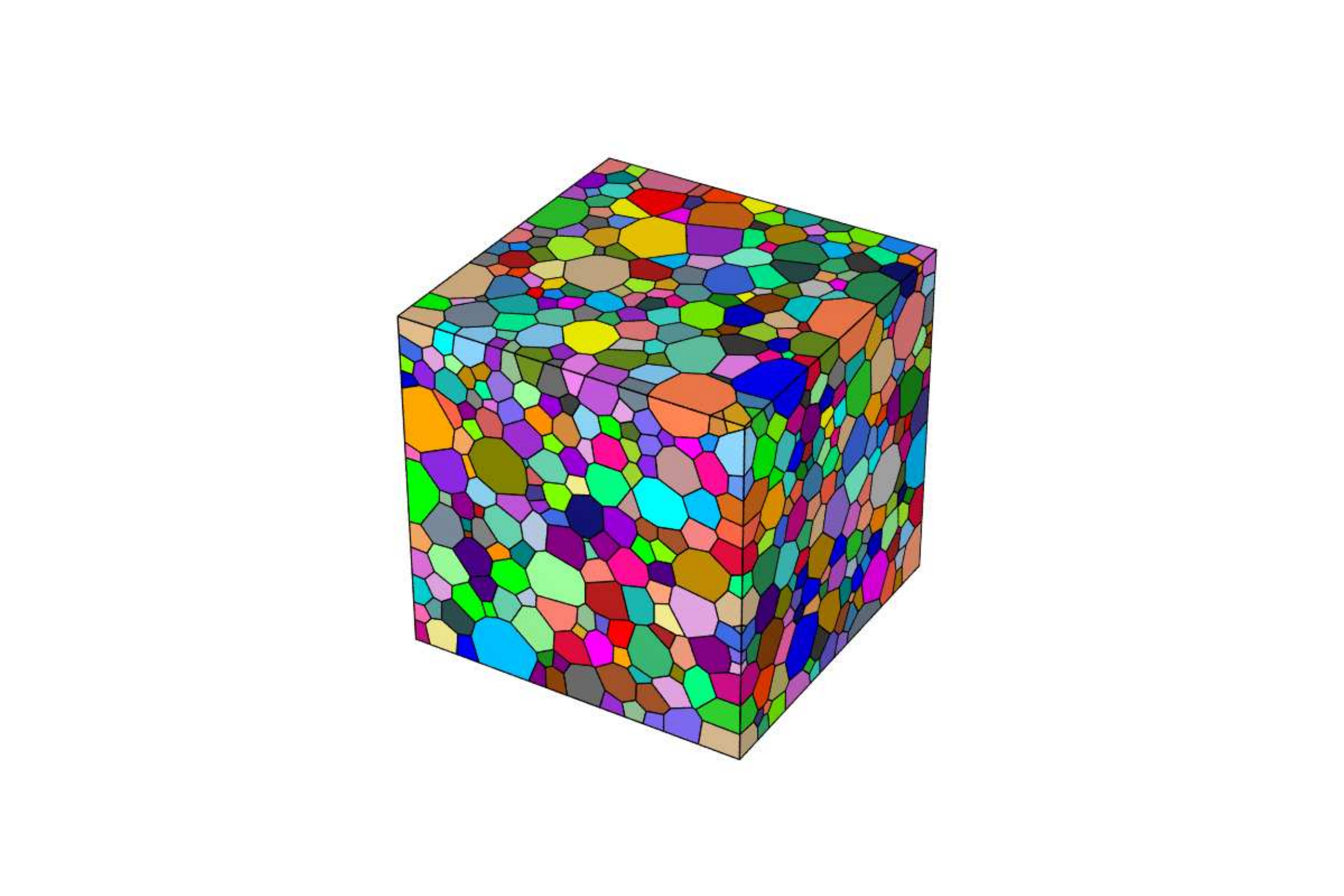}
    \caption{}
  \end{subfigure}
   \caption{Reconstructed cubical meshes of Estaillades carbonate}
  \label{fig:EC_model}
\end{figure}

\begin{figure}[!ht]
  \begin{subfigure}{.49\textwidth}
    \centering
    \hspace{0\textwidth}
    \includegraphics[scale=0.4, trim={2cm 9cm 2cm 9cm}, clip]{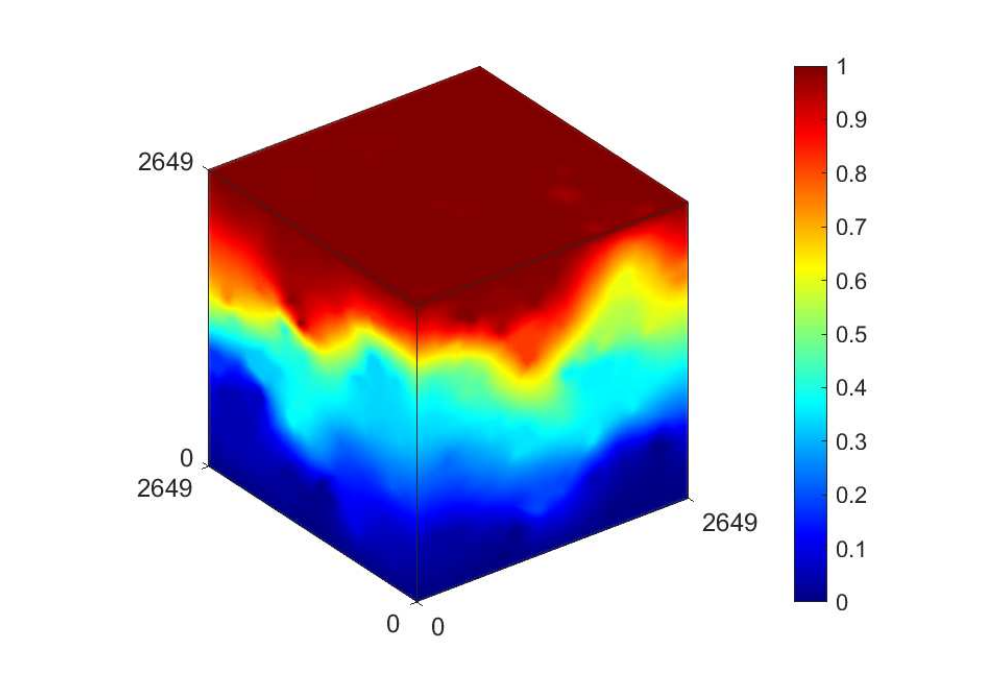}
    \caption{}
  \end{subfigure} 
  \begin{subfigure}{.49\textwidth}
    \centering
    \hspace{0\textwidth}
    \includegraphics[scale=0.4, trim={2cm 9cm 2cm 9cm}, clip]{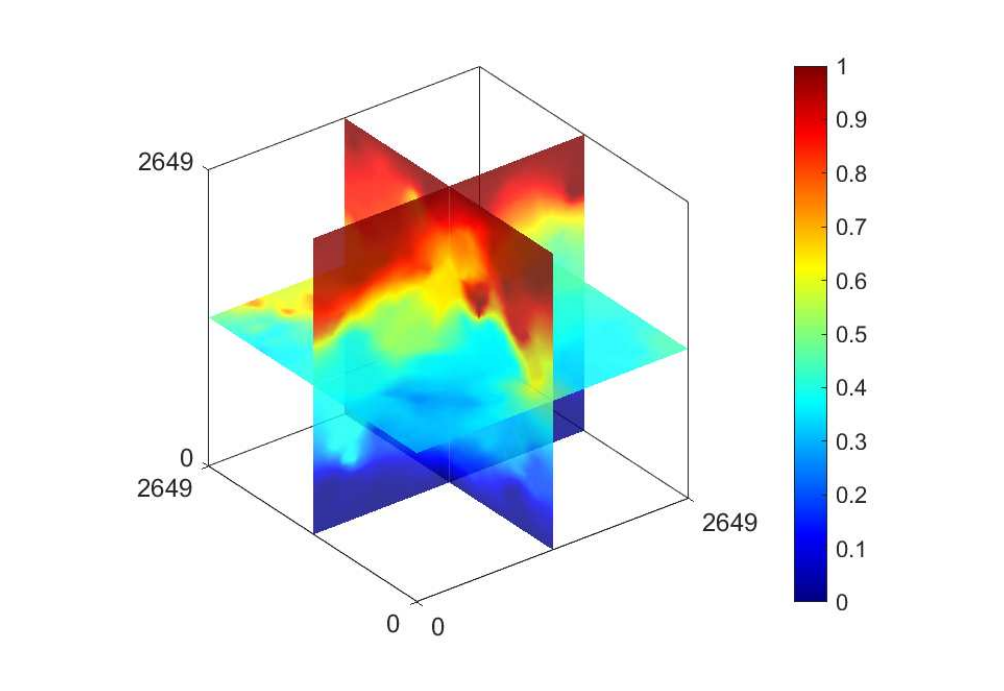}
    \caption{}
  \end{subfigure}
   \caption{Pressure ($\mathrm{Pa}$) distribution in Estaillades carbonate ($\mu m$)}
  \label{fig:EC_pressure}
\end{figure}

\begin{figure}[!ht]
  \begin{subfigure}{.49\textwidth}
    \centering
    \hspace{-0.1\textwidth}
    \includegraphics[scale=0.2, trim={2cm 2cm 2cm 2cm}, clip]{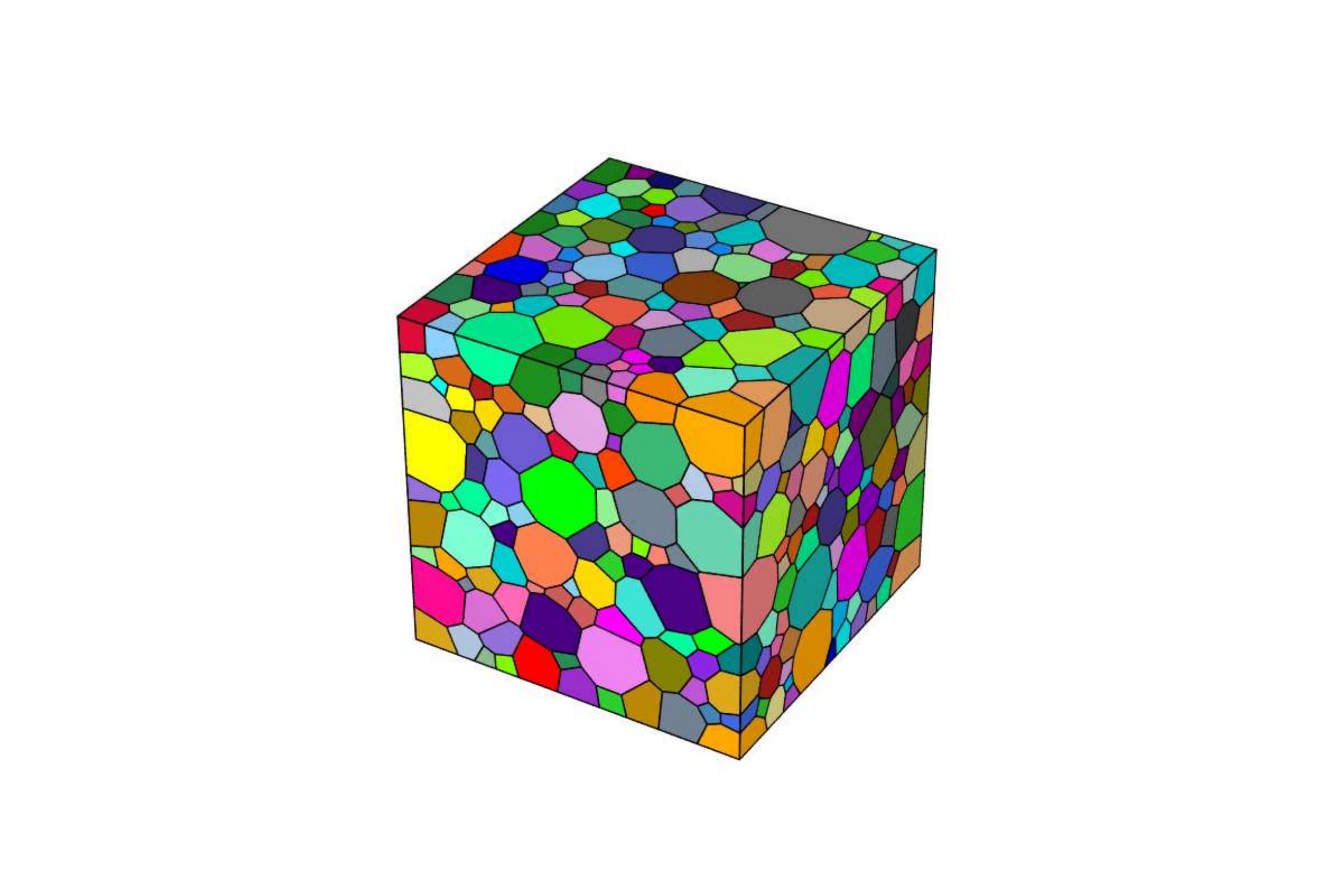}
    \caption{}
  \end{subfigure} 
  \begin{subfigure}{.49\textwidth}
    \centering
    \hspace{-0.1\textwidth}
    \includegraphics[scale=0.2, trim={2cm 2cm 2cm 2cm}, clip]{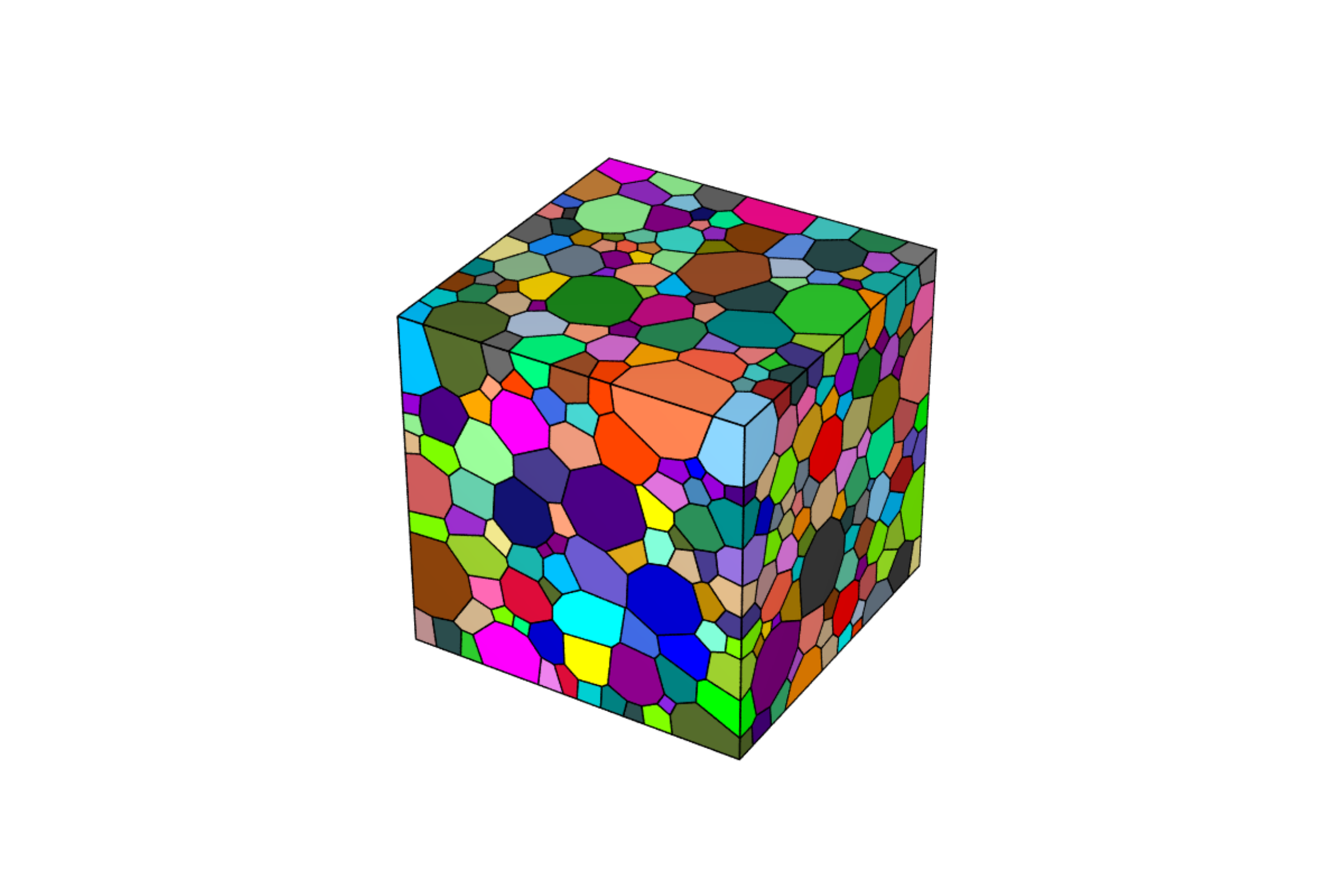}
    \caption{}
  \end{subfigure}
   \caption{Reconstructed cubical meshes of Ketton carbonate}
  \label{fig:KC_model}
\end{figure}

\begin{figure}[!ht]
  \begin{subfigure}{.49\textwidth}
    \centering
    \hspace{0\textwidth}
    \includegraphics[scale=0.4, trim={2cm 9cm 2cm 9cm}, clip]{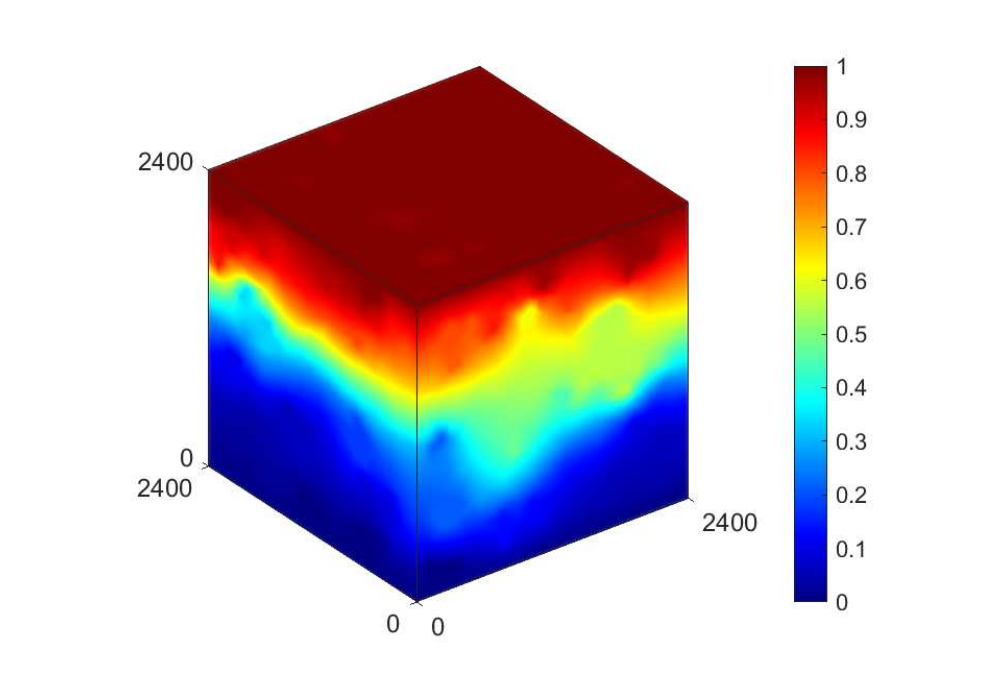}
    \caption{}
  \end{subfigure} 
  \begin{subfigure}{.49\textwidth}
    \centering
    \hspace{0\textwidth}
    \includegraphics[scale=0.4, trim={2cm 9cm 2cm 9cm}, clip]{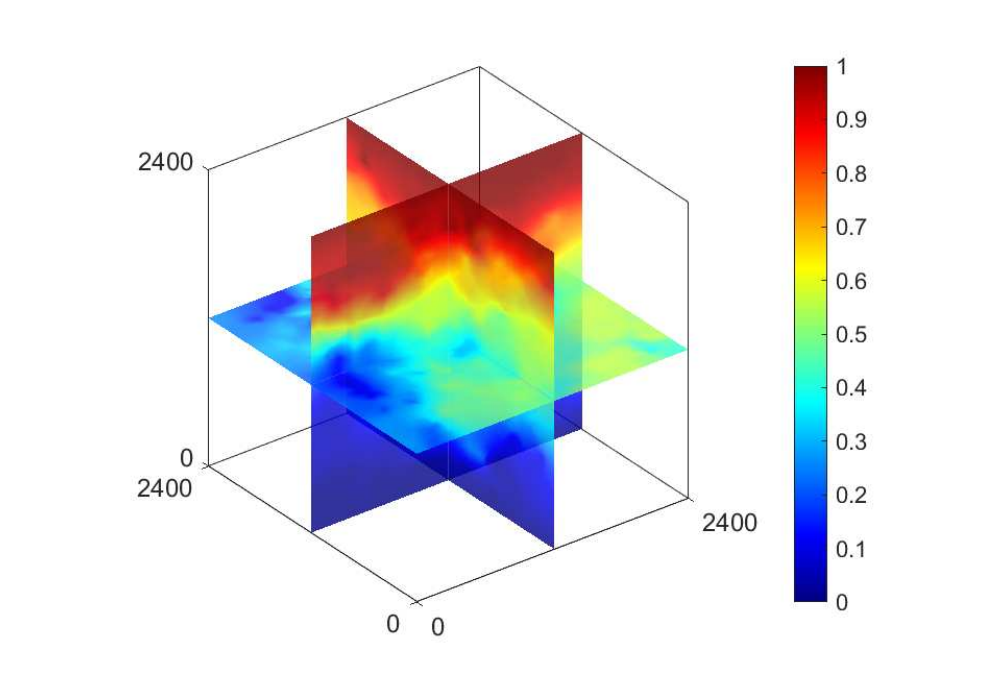}
    \caption{}
  \end{subfigure}
   \caption{Pressure distribution ($\mathrm{Pa}$) in Ketton carbonate ($\mu m$)}
  \label{fig:KC_pressure}
\end{figure}

\noindent Two model realisations and one calculated pressure distribution are shown in the folloing figures: Doddington sandstone (\Cref{fig:DS_model}, \Cref{fig:DS_pressure}), Estaillades carbonate (\Cref{fig:EC_model}, \Cref{fig:EC_pressure}) and Ketton carbonate (\Cref{fig:KC_model}, \Cref{fig:KC_pressure}).
\clearpage

\end{document}